%% file: L_HARQ.tex
\documentclass[12pt,draftcls,journal,letterpaper,twosides,onecolumn]{IEEEtran}

\usepackage{etex}

\input{Common.Files/included_packages.tex}
\usepackage[dvipsnames]{xcolor}
\input{Common.Files/commands.tex}

\usepackage[acronym,nonumberlist]{glossaries} 
\usepackage[draft]{hyperref}
\input{Common.Files/acronyms.tex}

\usepackage{tikz,pgfplots}
\pgfplotsset{width=3.4in,height=2.3in} 
\usetikzlibrary{positioning,shadows,shapes,fit,arrows,backgrounds}

\usepackage{epsf}           
\usepackage{graphicx}       
\usepackage{epsfig}         
\usepackage{pstricks}
\usepackage{pstricks-add}
\usepackage{pst-plot}
\usepackage{dsfont}

\usepackage{balance}

\newcommand{\siz}{1.2}  
\newcommand{\sizf}{0.65}   

\begin{document}


\title{How to Boost the Throughput of HARQ\\ with Off-the-Shelf Codes}

\author{
Mohammed Jabi, \'Etienne Pierre-Doray, Leszek Szczecinski, and Mustapha Benjillali
\\
\thanks{%
M.~Jabi and L.~Szczecinski are with INRS-EMT, Montreal, Canada. [e-mail: \{jabi,~leszek\}@emt.inrs.ca].}%
\thanks{%
\'E. Pierre-Doray is with Polytechnique de Montreal, Canada. He was with INRS-EMT when this work was carried out. [e-mail: etipdoray@gmail.com].}%
\thanks{%
M. Benjillali is with the Communication Systems Department, INPT, Rabat, Morocco. [e-mail: benjillali@ieee.org].}%
\thanks{%
Project partially financed by NSERC, Canada under ENGAGE grant EGP~490796--15.}%
}


\maketitle
\thispagestyle{empty}

\begin{abstract}
In this work, we propose  a coding strategy designed to enhance the throughput of \gls{harq} transmissions over i.i.d. block-fading channels with the \gls{csi} unknown at the transmitter. We use a joint packet coding where the same channel block is logically shared among many packets. To reduce the complexity, we use a two-layer coding where, first, packets are first coded by the binary compressing encoders, and the results are then passed to the conventional channel encoder. We show how to optimize the compression rates on the basis of the empirical error-rate curves. We also discuss how the parameters of the practical turbo-codes may be modified to take advantage of the proposed \gls{harq} scheme. Finally, simple and pragmatic rate adaptation strategies are developed.  In numerical examples, our scheme is compared to the conventional \gls{irharq}, and it yields a notable gain of $1-2\dBval$ in the region of high throughput, where \gls{harq} fails to provide any improvement. \end{abstract}


\input{Includes/Introduction.tex}

\input{Includes/Model.tex}
\input{Includes/L_HARQ_optimization.tex}

\input{Includes/Numerical.tex}
\input{Includes/Discussion.tex}


\section{Conclusions}\label{Sec:Conclusions}
In this work, we proposed an \gls{harq} transmission scheme and showed how its throughput can be optimized using  \gls{per} curves of the practical decoder. Compared to the conventional \gls{irharq} protocol, the proposed solution yields notable gains in the high throughput regime. In wireless systems, these gains may translate into energy savings, reduced intercell interference, or coverage extension. 

To illustrate our findings, we used turbo-codes to demonstrate the possibility of boosting \gls{harq} throughput with off-the-shelf codes, and we discussed the importance of a code design (here--the puncturing) to see the gains materialize. We only need the  simulated/measured \gls{per} curves $\PER(\SNR;\R)$ and $\PER(\SNR; \R, \Rs{})$ to perform the rate adaptation. Thus, our approach is well suited to the case of finite block-length, a promising feature for 5G systems which was studied recently in a similar context in \cite{Trillingsgaard14,Nguyen15}.  

Furthermore, we developed suboptimal but very simple rate adaptation strategies, and showed that the inflicted performance loss is negligible compared to the optimal schemes.



\balance

\bibliographystyle{IEEEtran}

\end{document}

%% file: Common.Files/included_packages.tex
\usepackage{amsmath,amssymb,amscd,latexsym,dsfont,wasysym,bbm}
\usepackage{float}
\usepackage{color}
\usepackage{graphicx,subfigure}
\usepackage{multirow,multicol}
\usepackage{verbatim}
\usepackage{psfrag}
\usepackage{comment}
\usepackage{makeidx}
\usepackage[]{algorithm}
\usepackage{algorithmic}
\usepackage{cite}
\usepackage{cases}
\usepackage{amsthm}
 \usepackage{setspace}

%% file: Common.Files/commands.tex
\newcommand{\tr}[1]{\textrm{#1}}
\newcommand{\mr}[1]{\mathrm{#1}}
\newcommand{\tnr}[1]{{\textnormal{#1}}}

\newcommand{\mc}[1]{\mathcal{#1}}
\newcommand{\mf}[1]{\mathsf{#1}}

\newcommand{\ms}[1]{\mathds{#1}}

\newcommand{\ov}[1]{\overline{#1}}

\newcommand{\bx}{\boldsymbol{x}}

\newcommand{\by}{\boldsymbol{y}}

\newcommand{\bz}{\boldsymbol{z}}

\newcommand{\figref}[1]{Fig.~\ref{#1}}

\newcommand{\secref}[1]{Sec.~\ref{#1}}

\newcommand{\ie}{i.e.,~} 		
\newcommand{\eg}{e.g.,~}	

\newcommand{\argmax}{\mathop{\mr{argmax}}}
\newcommand{\argmin}{\mathop{\mr{argmin}}}

\newcommand{\set}[1]{\{#1\}}
\newcommand{\SET}[1]{\left\{#1\right\}}

\newcommand{\cd}{\cdot}
\newcommand{\ld}{\ldots}

\newcommand{\PR}[1]{\Pr\SET{#1}}       	
\newcommand{\pdf}{p}            			

\newcommand{\IND}[1]{\ms{I}\big[{#1}\big]}   	
\newcommand{\Ex}{\ms{E}}     			

\newcommand{\mcA}{\mc{A}}

\newcommand{\mcF}{\mc{F}}

\newcommand{\mcX}{\mc{X}}

\newcommand{\mfc}{\mf{c}}

\newcommand{\mfm}{\mf{m}}

\newcommand{\mfR}{\mf{R}}
\newcommand{\mfD}{\mf{D}}

\newcommand{\Real}{\mathbb{R}}		

\newcommand{\SNR}{\mathsf{snr}}  
\newcommand{\SNRrv}{\mathsf{SNR}}  
\newcommand{\SNRav}{\ov{\mathsf{snr}}}  
\newcommand{\X}{\mcX}	

\newcommand{\R}{R}              	
\newcommand{\Ns}{{\mathop{N_\tnr{s}}}} 	
\newcommand{\Nc}{N_\tnr{c}} 	

\newcommand{\IR}{\tnr{ir}}

\newcommand{\Err}{\mf{ERR}}
\newcommand{\Errb}[1]{\mf{ERR}^{\tnr{b}}_{#1}}       
\newcommand{\nack}{\mf{NACK}}  
\newcommand{\kmax}{K}		

\newcommand{\Lc}{\tr{L}} 
\newcommand{\AN}{\tr{AoN}} 
\newcommand{\xp}{\tnr{xp}}

\newcommand{\TmR}{T_{R} } 
\newcommand{\Rs}[1]{\mathop{\rho_{#1}}}       

\newcommand{\PER}{\mr{PER}}  

\newcommand{\dBval}{\tnr{dB}} 
\newcommand{\dB}{~\tnr{[dB]}} 

%% file: Common.Files/acronyms.tex
\newacronym[\glsshortpluralkey=PDFs,\glslongpluralkey=probability density functions]{pdf}{PDF}{probability density function}
\newacronym[\glsshortpluralkey=CDFs,\glslongpluralkey=cumulative density functions]{cdf}{CDF}{cumulative density function}
\newacronym[\glsshortpluralkey=CCDFs,\glslongpluralkey=complementary cumulative density functions]{ccdf}{CDF}{complementary cumulative density function}
\newacronym[\glsshortpluralkey=PMFs,\glslongpluralkey=probability mass functions]{pmf}{PMF}{probability mass function}
\newacronym[]{lhs}{l.h.s.}{left-hand side}
\newacronym[]{rhs}{r.h.s.}{right-hand side} 

\newacronym[]{bicm}{BICM}{bit-interleaved coded modulation}
\newacronym[]{bicmid}{BICM-ID}{BICM with iterative demapping}
\newacronym[]{cm}{CM}{coded modulation}
\newacronym[]{tcm}{TCM}{trellis-coded modulation}
\newacronym[]{mlc}{MLC}{multi-level coding}
\newacronym[]{pam}{PAM}{pulse amplitude modulation}
\newacronym[]{bpsk}{BPSK}{binary phase shift keying}
\newacronym[]{qam}{QAM}{quadrature amplitude modulation}
\newacronym[]{psk}{PSK}{phase shift keying}
\newacronym[\glsshortpluralkey=LLRs,\glslongpluralkey=logarithmic likelihood ratios]{llr}{LLR}{logarithmic likelihood ratio}
\newacronym[]{map}{MAP}{maximum a posteriori}
\newacronym[]{ml}{ML}{maximum likelihood}
\newacronym[\glsshortpluralkey=MIs,\glslongpluralkey=mutual informations]{mi}{MI}{mutual information}
\newacronym[\glsshortpluralkey=GMIs,\glslongpluralkey=generalized mutual informations]{gmi}{GMI}{generalized mutual information}
\newacronym[]{eesm}{EESM}{exponential effective-SNR-mapping}
\newacronym[]{bicm-gmi}{BICM-GMI}{BICM generalized mutual information}
\newacronym[]{awgn}{AWGN}{additive white Gaussian noise}
\newacronym[]{amc}{AMC}{adaptive modulation and coding}
\newacronym[]{csi}{CSI}{channel state information}
\newacronym[]{cqi}{CQI}{channel quality indicator}
\newacronym[]{sp}{SP}{set-partitioning}
\newacronym[]{gsm}{GSM}{global system for mobile communications}
\newacronym[]{edge}{EDGE}{enhanced data rates for GSM evolution}
\newacronym[]{3gpp}{3GPP}{3rd generation partnership project}
\newacronym[]{lte}{LTE}{Long Term Evolution}
\newacronym[]{dvb}{DVB}{digital video broadcasting}

\newacronym[\glsshortpluralkey=CCs,\glslongpluralkey=convolutional codes]{cc}{CC}{convolutional code}
\newacronym[\glsshortpluralkey=PCCCs,\glslongpluralkey=parallel concatenated convolutional codes]{pccc}{PCCC}{parallel concatenated convolutional code}
\newacronym[\glsshortpluralkey=TCs,\glslongpluralkey=turbo codes]{tc}{TC}{turbo code}
\newacronym{ldpc}{LDPC}{low-density parity-check}
\newacronym[]{ofdm}{OFDM}{orthogonal frequency-division multiplexing}
\newacronym[]{bep}{BEP}{bit-error probability}
\newacronym[]{wep}{WEP}{word-error probability}
\newacronym[]{sep}{SEP}{symbol-error probability}
\newacronym[]{pep}{PEP}{pairwise-error probability}
\newacronym[]{ttcm}{TTCM}{turbo-trellis coded modulation}
\newacronym[]{uep}{UEP}{unequal error protection}
\newacronym[\glsshortpluralkey=CENCs,\glslongpluralkey=convolutional encoders]{cenc}{CENC}{convolutional encoder}
\newacronym[]{mimo}{MIMO}{multiple-input multiple-output}
\newacronym[\glsshortpluralkey=SNRs,\glslongpluralkey=signal-to-noise ratios]{snr}{SNR}{signal-to-noise ratio}
\newacronym[\glsshortpluralkey=SINRs,\glslongpluralkey=the signal-to-interference-plus-noise ratios]{sinr}{SINR}{the signal-to-interference-plus-noise ratio}
\newacronym[]{msb}{MSB}{most-significative bit}
\newacronym[]{bcjr}{BCJR}{Bahl--Cocke--Jelinek--Raviv}
\newacronym[\glsshortpluralkey=SEDs,\glslongpluralkey=squared Euclidean distances]{sed}{SED}{squared Euclidean distance}
\newacronym[\glsshortpluralkey=EDs,\glslongpluralkey=Euclidean distances]{ed}{ED}{Euclidean distance}
\newacronym[\glsshortpluralkey=MEDs,\glslongpluralkey=minimum Euclidean distances]{med}{MED}{minimum Euclidean distance}
\newacronym[]{core}{CoRe}{constellation rearrangement}
\newacronym[]{msd}{MSD}{multistage decoding}
\newacronym[]{pdl}{PDL}{parallel decoding of the individual levels}
\newacronym[\glsshortpluralkey=GCs,\glslongpluralkey=Gray codes]{gc}{GC}{Gray code}
\newacronym[]{brgc}{BRGC}{binary-reflected Gray code}
\newacronym[]{nbc}{NBC}{natural binary code}
\newacronym[]{fbc}{FBC}{folded-binary code}
\newacronym[]{bsgc}{BSGC}{binary semi-Gray code}
\newacronym[]{msp}{MSP}{modified set-partitioning}
\newacronym[]{ssp}{SSP}{semi set-partitioning}
\newacronym[]{fhd}{FHD}{free Hamming distance}
\newacronym[]{mfhd}{MFHD}{maximum free Hamming distance}
\newacronym[]{ods}{ODS}{optimal distance spectrum}
\newacronym[]{iud}{i.u.d.}{independent and uniformly distributed}
\newacronym[]{ud}{u.d.}{uniformly distributed}
\newacronym[]{iid}{i.i.d.}{independent, identically distributed}
\newacronym[]{bico}{BICO}{binary-input continuous-output}
\newacronym[]{gh}{GH}{Gauss--Hermite}

\newacronym[\glsshortpluralkey=BSs,\glslongpluralkey=base-stations]{bs}{BS}{base-station}
\newacronym[\glsshortpluralkey=MSs,\glslongpluralkey=mobile-stations]{ms}{MS}{mobile-stations}

\newacronym[]{phy}{PHY}{physical layer} 
\newacronym[]{llc}{LLC}{logical link control} 
\newacronym[]{mac}{MAC}{media access control} 
\newacronym[]{fft}{FFT}{fast Fourier transform} 
\newacronym[]{cf}{CF}{characteristic function} 
\newacronym[]{mgf}{MGF}{moment generating function} 
\newacronym[]{ee}{EE}{energy efficiency} 
\newacronym[]{kkt}{KKT}{Karush--Kuhn--Tucker} 
\newacronym[]{mcs}{MCS}{modulation/coding scheme} 
\newacronym[]{fec}{FEC}{forward error correction}
\newacronym[]{arq}{ARQ}{automatic repeat request}
\newacronym[]{harq}{HARQ}{hybrid ARQ}
\newacronym[]{tarq}{TARQ}{truncated HARQ}
\newacronym[]{rrharq}{RR-HARQ}{repetition redundancy HARQ}
\newacronym[]{irharq}{IR-HARQ}{incremental redundancy HARQ}
\newacronym[]{ack}{ACK}{positive acknowledgment}
\newacronym[]{nack}{NACK}{negative acknowledgment}
\newacronym[]{crc}{CRC}{cyclic redundancy check}
\newacronym[]{dp}{DP}{dynamic programming}
\newacronym[]{gp}{GP}{geometric programming}
\newacronym[]{per}{PER}{packet error rate}
\newacronym[]{op}{OP}{outage probability}
\newacronym[]{spa}{SPA}{saddle-point approximation}
\newacronym[]{mrc}{MRC}{maximum ratio combining}
\newacronym[]{mdp}{MDP}{Markov decision process}
\newacronym[]{pomdp}{POMDP}{Partially observable MDP}
\newacronym[]{psimdp}{PSI-MDP}{Partial State Information Markov Decision Process}

\newacronym[]{mm}{MM-HARQ}{multi-message HARQ}
\newacronym[]{xp}{XP-HARQ}{Cross-packet HARQ}
\newacronym[]{ts}{TS}{time-sharing}
\newacronym[]{sc}{SC}{superposition coding}
\newacronym[]{sbrq}{SBRQ}{systematic backward retransmission}
\newacronym[]{brq}{BRQ}{backward retransmission}
\newacronym[]{lharq}{L-HARQ}{layer-coded HARQ}
\newacronym[]{anlharq}{AoN-HARQ}{all-or-none L-HARQ}

\newacronym[]{pp}{PPP}{point process}
\newacronym[]{ppp}{PPP}{Poisson point process}
\newacronym[]{pgfl}{PGFL}{Poisson point process}

%% file: Includes/Introduction.tex
\section{Introduction}\label{Sec:Intro}
In this work, we propose and analyze a Hybrid ARQ protocol  based on practical (``off-the-shelf'') codes whose parameters are optimized to maximize the throughput for transmission over block-fading channels.%

\gls{harq} protocols are used to guarantee a reliable communication over error-prone channels, where the receiver uses the feedback to inform the transmitter about the decoding success (via \gls{ack} messages) or failure (via \gls{nack} messages). After each \gls{nack}, the transmitter starts a new \gls{harq} round (or, a \emph{retransmission}); this continues till the \gls{ack} message is received or the maximum allowed number of rounds is attained.%

In this work, we assume that the transmitter operates without the instantaneous \gls{csi}, so the retransmissions in \gls{harq} can be considered as an implicit adaptation to the channel states: each \gls{nack} triggers the transmission of additional parts of the codewords, and hence reduces the effective coding rate which in turn facilitates the decoding of the packet. Such a setup became ``canonical'' with the work \cite{Caire01} which demonstrated that the throughput of \gls{harq} can approach the ergodic capacity, and this, despite a binary and per-block feedback. However, to attain the ergodic capacity, \cite{Caire01} assumes a very  high coding rate per round, $\R$, and  a very large number of transmission rounds; since large memories at the transmitter and the receiver are then necessary,  this approach is impractical.%

The practical problem is thus to increase the throughput for a \emph{given} and \emph{finite} rate $\R$. This problem is particularly challenging for the throughput in the vicinity of $\R$, where the conventional \gls{harq} fails to provide any improvement \cite{Larsson14,Jabi15b}.%

To address this issue, two main venues have been explored in the literature. The first relies on the explicit reduction of the required transmission time, see \eg \cite{Cheng03,Uhlemann03,Visotsky03,Visotsky05,Pfletschinger10,Szczecinski13}. However, the  throughput increase is obtained with variable-length channel blocks which may be a challenge in those systems which have to keep the block size constant. The second venue harnesses the channel coding to overcome this very difficulty: the works \cite{Larsson13,Popovski14,Trillingsgaard14,Nguyen15,Jabi15b,Benyouss16} keep the block size constant but increase the coding rate, \ie the number of bits encoded in each \gls{harq} round. This may be seen as a joint encoding of various packets into a single channel block. Then, the challenge is to define a simple (joint) encoding/decoding strategy and to optimize the coding rates.%

In this work, we pursue the second venue with two main objectives, namely 1)~To use off-the-shelf encoders and decoders, and 2)~To optimize the transmission parameters (rates) of truncated \gls{harq}. In fact, both objectives are interconnected since the ``off-the-shelf'' (\ie simple to implement) encoders/decoders must also be accompanied by simple tools allowing us to optimize the coding rates; more on that in \secref{Sec:MM.HARQ}.%


The contributions of this work are the following:
\begin{itemize}
\item We compare the implementation feasibility of various joint coding strategies in the light of the implementation/optimization simplicity and we propose to use \gls{lharq} which is a modified version of  \gls{harq} proposed in~\cite{Popovski14}.
\item We show how to calculate the throughput of  truncated \gls{lharq} based on the off-the-shelf encoders/decoders. Our approach is applicable to any scenario where the empirical error-rate curves characterizing the decoders are known. This is different from~\cite{Popovski14} which assumed an infinite number of rounds and an idealized coding/decoding.%
\item We formulate and solve the problem of rate adaptation using a \gls{dp}  and compare the throughputs of \gls{lharq} to those of conventional \gls{irharq}. While \cite{Nguyen15,Benyouss16} addressed the issue of rate optimization for idealized-decoding scenarios and explicitly joint (\ie non layer) decoding, to the best of our knowledge, none of the previous works addressed the issue of rate optimization with off-the-shelf encoders/decoders.
\item We show the throughput achievable with \gls{lharq} based on (turbo)-codes, where the optimal solution is found using solely the empirical error-rate curves of the decoder. We also discuss the issue of choosing the encoder parameters (puncturing pattern) and its relationship with the performance of \gls{lharq}.%
\item We propose and optimize  a simplified version of \gls{lharq}.
\end{itemize}

The rest of the paper is organized as follows. We define the system model and introduce the considered retransmission schemes in~\secref{Sec:Model}. The proposed layer-coded \gls{harq} is defined in~\secref{sec:sbrq}, the rate optimization procedure is explained in~\secref{Sec:multibit} and illustrated with numerical results shown in~\secref{Sec:Numerical}. Next, we discuss the sub-optimal rate adaptation policies in~\secref{sec:discussion}. Conclusions are drawn in~\secref{Sec:Conclusions}.

%% file: Includes/Model.tex
\section{Incremental Redundancy HARQ}\label{Sec:Model}

In conventional \gls{irharq}, a packet $\mfm \in\set{0,1}^{\R\Ns}$ is encoded into $\kmax$ subcodewords $\bx_k=\Phi_k[\mfm]\in\X^{\Ns}$, each composed of $\Ns$ complex symbols drawn from a constellation $\X$, where $\Phi_k[\cd]$ are the encoders generating complementary/incremental redundancy symbols; here $\R$ denotes the coding rate per block.\footnote{As the number of used subcodewords is random, we find it more convenient to define the rate per channel block (or per subcodeword), instead of the rate per the entire codeword $\R/\kmax$ because transmission with such a rate is a random event.}

We consider a point-to-point transmission over a block fading channel. Each packet may require many transmission \emph{rounds}. The $k$th round carries a subcodeword $\bx_k$ and the received signal is given by
\begin{align}\label{y.x.z}
\by_{k}=\sqrt{\SNR_{k}}\bx_{k}+\bz_{k},\quad k=1,\ld, \kmax,
\end{align}
where $\bz_{k}$ is a zero mean, unit-variance, complex Gaussian variable modeling the noise, $\kmax$ is the maximum number of rounds; fixing the average energy of $\bx_k$ to unity, and $\SNR_{k}$ is the \gls{snr} at the receiver, which we assume to be perfectly known/estimated at the receiver but unknown at the transmitter. 

We will model $\SNR_{k}$ by \gls{iid} random variables $\SNRrv_k$. The derivations will be done in abstraction of a particular fading type, but in the numerical examples we consider the Rayleigh fading model, hence, $\SNRrv_k$ follow exponential distributions
\begin{align}
\pdf_{\SNRrv_k}(\SNR)=\frac{1}{\SNRav}\exp(-\SNR/\SNRav),
\end{align}
where $\SNRav$ is the average \gls{snr}.%

After the transmission in the $k$th round, the receiver tries to decode the packet $\mfm$ using all the received channel outcomes
\begin{align}\label{by}
\hat{\mfm}_k&=\tnr{DEC}[\by_{1},\ld,\by_{k-1},\by_{k} ],
\end{align}
and, using a binary feedback channel, informs the transmitter whether the decoding succeeded, \ie $\set{\hat{\mfm}_k=\mfm}$ (through an \gls{ack}) or failed (through a \gls{nack}). The transmission rounds continue until an \gls{ack} is received or the $\kmax$th round is reached. 

\subsection{Throughput}\label{Sec:HARQIR.Throughput}
The \gls{harq} \emph{cycle} is a sequence of $\mfD$ transmission rounds related to the same packet $\mfm$. In truncated \gls{harq}, $\mfD\leq\kmax$. Each round may be seen as a state of a Markov chain. At the end of the cycle (the ``renewal", in the language of Markov processes), the receiver obtains a ``reward" $\mfR\in\set{0,\R}$, which is the number of correctly received bits normalized by the number of symbols in the block, $\Ns$. 

Since $\mfD$ and $\mfR$ are random, the long-term average throughput is calculated from the reward-renewal theorem, as the ratio between the expected reward and the expected duration \cite{Caire01}, 
\begin{align}\label{eta.C.IR}
\eta_{\kmax}^{\IR}
&=\frac{\Ex[\mfR]}{\Ex[\mfD]}=\frac{\R(1-f_{\kmax})}{\sum_{k=0}^{\kmax-1}f_{k}},
\end{align}
which we specialized for the case of truncated \gls{harq} \cite[Sec.~III]{Jabi16} using the probability of the decoding failure after $k$ rounds
 \begin{align}\label{f.k}
f_{k}=\Pr\set{\nack_k},
\end{align}
where
 \begin{align}\label{}
\nack_k&\triangleq\Bigl\{ \Err_1 \wedge \Err_2 \wedge \ld\wedge \Err_k \Bigr\}
\end{align}
and $\Err_k\triangleq\set{\hat{\mfm}_k\neq\mfm}$ denotes the event of a decoding error in the $k$th round.

Therefore, to evaluate the throughput, which is our metric of interest, we need to calculate $f_{k}$. 

In the idealized model of \cite{Caire01,Larsson14,Jabi16}, it is assumed that $\Err_k=\set{\sum_{l=1}^k I(\SNR_l)<\R}$, where  $I(\SNR_k)$ is the \gls{mi} between the channel input and output in the $k$th block; then, $\nack_k\iff\Err_k$ is deterministically defined by the values of the \gls{snr}s. 

In practice,  however, the decoding errors depend also on the information sequence and the realizations of the noise. The expectation taken with respect to these variables  yields the \gls{per} curve of the decoder, 
\begin{align}\label{Per:MD}
\PER(\SNR_1,\ld,\SNR_k;\R)\triangleq\Pr\set{\Err_k|\SNR_1,\ld,\SNR_k,\R},
\end{align}
which may be obtained with Monte-Carlo simulations, keeping the \gls{snr}s and the transmission rate $\R$ fixed.  

Under such a model, the events $\Err_k$ and $\nack_k$ are not identical. Nevertheless, we may use the approximate relation  of backward decoding error implication $\Err_k\implies\Err_{k-1}\implies\ld\implies\Err_{1}$ \cite{Gu06,Long09}, which allows to write $\PR{\nack_k}\approx\PR{\Err_k}$.

\subsection{Cross-packet coding for HARQ}\label{Sec:MM.HARQ}

As observed before, \eg in \cite{Larsson14,Jabi15b,Jabi16}, \gls{harq} is particularly useful when the probability of error in the first round $f_1$ is high, as then the throughput can be notably increased with $\kmax$. On the other hand, \gls{harq} has negligible impact on the throughput when $f_1 \ll 1$; this is because $f_k<f_{1}^k \ll f_1$, and then 
\begin{align}\nonumber
\eta_{\kmax}^{\IR}=\frac{\R(1-f_{\kmax})}{1+f_1+\sum_{k=2}^{\kmax-1}f_k}\approx\frac{\R}{1+f_1}\approx\R(1-f_1)=\eta_1,
\end{align}
where $\eta_1$ is the throughput of one-round (non-\gls{harq}) transmission. Thus, we cannot expect any improvement in the throughput deploying conventional \gls{irharq} for relatively small $f_1$, or---alternatively---for $\eta_1$ close to $\R$ \cite{Larsson14,Jabi15b,Jabi16}. In our model it also means that \gls{irharq} is not useful for high average \gls{snr}. 

\begin{figure}[bt]
\begin{center}
\scalebox{0.65}{\input{figures/MM_model_V4}}
\end{center}
\caption{Model of the joint coding/decoding \gls{harq} transmission. The \gls{harq} controller has to adjust the coding rates using feedback information.}\label{Fig:XP.HARQ}
\end{figure}
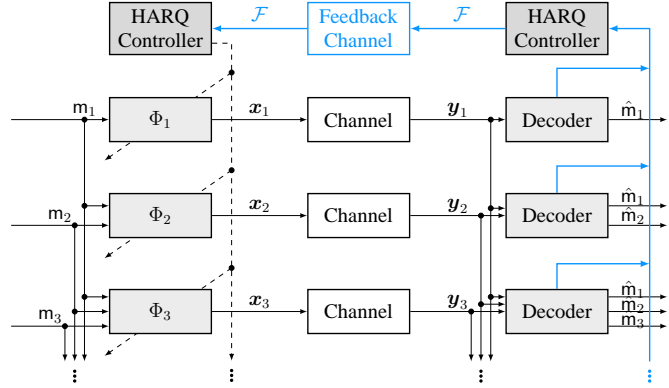

The reason is that, due to predefined coding, the reward $\mfR$ is not allowed to grow even if $\mfD$ increases throughout the \gls{harq} rounds. Thus, to improve the throughput, the coding should be modified so as to increase the attainable reward as the rounds advance. To this end we let the transmitter to jointly encode multiple packets into the same codeword as shown in~\figref{Fig:XP.HARQ}
\begin{align}\label{xp.coding}
\bx_k&=\Phi_k[\mfm_{[k]}]\in\mcX^\Ns\\
\mfm_{[k]}&=[\mfm_1,\ld,\mfm_k]\in\set{0,1}^{\Ns\R_{[k]}},
\end{align}
where $\R_{[k]}$ denotes the joint coding rate in the $k$th round. The throughput of such \gls{xp} is calculated as \cite{Benyouss16}
\begin{align}\label{eta.C.IR.XP}
\eta_{\kmax}^{\xp}
&=\frac{\sum_{k=0}^{\kmax-1}\R_{[k]}(f_{k-1}-f_{k})}{\sum_{k=0}^{\kmax-1}f_{k}},
\end{align}
where $f_k$ is defined by \eqref{f.k} with $\Err_k=\set{\hat{\mfm}_{[k]}\neq\mfm_{[k]}}$ being the error of the  joint packet decoding, \ie
\begin{align}\label{joint.direct.decoding}
\hat{\mfm}_{[k]}=\tnr{DEC}[\by_1,\ld,\by_k].
\end{align}

Comparing to \eqref{eta.C.IR}, the throughput can be increased by increasing the numerator of \eqref{eta.C.IR.XP} if values of $\R_{[k]}$ are optimized.

To attain \eqref{eta.C.IR.XP} two main venues are adopted in the literature: i)~\emph{direct} encoding/decoding \cite{Trillingsgaard14,Nguyen15,Jabi15b,Benyouss16}, and ii)~\emph{layer} encoding/decoding \cite{Larsson13,Popovski14,Jabi15b}, which have different impact on the encoding/decoding complexity. 

The direct encoding considers \eqref{xp.coding} without any constraints on $\Phi_k[\cd]$; it is thus entirely general but raises some practical concerns regarding its implementation. Namely
\begin{enumerate}
\item The encoder $\Phi_k$ must accept inputs $\mfm_{[k]}$ with increasing lengths, $\Ns\R_1<\Ns\R_{[2]}<\ld<\Ns\R_{[k]}$,   while practical encoders are limited with regard to the input length (\eg due to the available encoding matrix in the \gls{ldpc} codes or the way the interleavers are defined in turbo-codes); 
\item Since the coding rates $\R_{[k]}$ grow with $k$ (and may even exceed $|\mcX|$), the customized design of the encoder $\Phi_k[\cd]$ is necessary to take into account the encoders used in the previous rounds $\Phi_l[\cd], l=1,\ld,k-1$. 
\item The joint decoding \eqref{joint.direct.decoding} must consider concatenation of the decoders and has implementation issues of its own as can be seen, for example, in \cite{Hausl07,Duyck10}.
\item The multi-dimensional \gls{per} curves \eqref{Per:MD}, depending on the coding rates, $\R_{[k]}$, would be very cumbersome to measure and store.  
\end{enumerate}

These issues make the direct encoding unfit to be used with ``off-the-shelf'' codes and thus, we will not follow this approach. Instead, we address the practical aspects with the layer-coded \gls{harq} (\gls{lharq}) we explain in the following.

%% file: figures/MM_model_V4.tex

\pgfdeclarelayer{background}
\pgfdeclarelayer{foreground}
\pgfsetlayers{background,main,foreground}

\definecolor{LHCblue}{RGB}{10, 150, 255}
\tikzstyle{form0} = [draw, minimum width=0cm, text width=0cm, 
  text centered,  minimum height=0cm]

\tikzstyle{form1} = [draw,  thick, minimum width=2cm, text width=1.8cm, fill=white, 
  text centered,  minimum height=.9cm]

\tikzstyle{form1b} = [color=LHCblue,draw,  thick, minimum width=2cm, text width=1.8cm, fill=white, 
  text centered,  minimum height=.9cm]

\tikzstyle{form1c} = [draw,  thick, minimum width=2cm, text width=1.8cm, fill=gray!15, 
  text centered,  minimum height=.9cm]

\tikzstyle{form1d} = [draw,  thick, minimum width=2cm, text width=1.8cm, fill=gray!30, 
  text centered,  minimum height=.9cm]

\tikzstyle{form2} = [draw=none, minimum width=3.4cm, text width=1.5cm, fill=none, 
  text centered,  minimum height=0cm]    

\tikzstyle{form3} = [color=LHCblue,draw, thick,minimum width=1.6cm, text width=1.5cm, 
  text centered,  minimum height=1.6cm]
  
  \tikzstyle{form4} = [draw=none, minimum width=0cm, text width=0cm, fill=none, 
  text centered,  minimum height=.6cm] 
  
    \tikzstyle{form5} = [draw, minimum width=1cm, text width=1cm,  fill=red!20,
text centered,  minimum height=1cm]

    \tikzstyle{form7} = [draw, minimum width=0.8cm, text width=0.8cm,fill=blue!20, 
text centered,  minimum height=0.6cm]

     \tikzstyle{form6} = [draw, minimum width=2cm, text width=.4cm, 
  text centered,  minimum height=1.5cm]
  
      \tikzstyle{form8} = [draw, minimum width=1.8cm, text width=1.8cm, fill=green!50!yellow!15,
  text centered,  minimum height=2.5cm]
  
   \tikzstyle{form9} = [draw, minimum width=2.5cm, text width=3cm, fill=green!50!yellow!15,
  text centered,  minimum height=4cm]
  
    \tikzstyle{form10} = [draw, minimum width=0.8cm, text width=0.8cm,fill=none, 
text centered,  minimum height=0.6cm]

\begin{tikzpicture}[trim left=-4.5cm]

    \node(Chan)[form1] { Channel};    
    
    \path (Chan.north)+(0,1.5)  node[form1](Chan1) {Channel}; 
     
     \path (Chan.south)+(0,-1.5)  node[form1](Chan3) {Channel}; 
            
     \path (Chan.west)+(-3,0)  node[form1c](enc2) {$\Phi_2$}; 
     \path (Chan.east)+(3,0)  node[form1c](dec2) {Decoder}; 
         
          \path (Chan1.west)+(-3,0)  node[form1c](enc1) {$\Phi_1$}; 
     \path (Chan1.east)+(3,0)  node[form1c](dec1) {Decoder}; 

        \path (Chan3.west)+(-3,0)  node[form1c](enc3) {$\Phi_3$}; 
     \path (Chan3.east)+(3,0)  node[form1c](dec3) {Decoder}; 

    \path (enc1.north)+(0,+1.4)  node[form1d](harq1) {HARQ Controller}; 
      \path (dec1.north)+(0,+1.4)  node[form1d](harq2) {HARQ Controller}; 
      
        \path (Chan1.north)+(0,+1.4)  node[form1b](feed) {Feedback Channel}; 
        
        
         
           \path (harq1.east)+(0,-0.3)  node(P1_harq1) {};  
             \path (P1_harq1.base)+(0.4,-6.5)  node(inf) {}; 
             
              \path (enc1.west)+(-2,0)  node(P1_m1) {}; 
              
               \path (enc2.west)+(0,0.2)  node(P1_enc2) {};
               
               \path (enc2.west)+(0,-0.2)  node(P2_enc2) {};
               
                \path (P1_enc2.base)+(-0.5,0)  node(P1_m2) {};  
                \path (P2_enc2.base)+(-2,0)  node(P2_m2) {}; 
                \path (P2_enc2.base)+(-0.7,0)  node(P2ref_m2) {};

                 \path (enc3.west)+(0,0.3)  node(P1_enc3) {};
                
                 \path (enc3.west)+(0,-0.3)  node(P3_enc3) {};

                   \path (P1_enc3.base)+(-0.5,0)  node(P1_m3) {}; 
                   \path (enc3.west)+(-0.7,0)  node(P2_m3) {};
                 \path (P3_enc3.base)+(-2,0)  node(P3_m3) {}; 
                  \path (P3_enc3.base)+(-0.9,0)  node(P3ref_m3) {};

                 \path (enc1.west)+(-0.5,0)  node(P1ref_m1) {};  
                  
                   \path (P1ref_m1.base)+(0,-4.95)  node(m1_inf) {}; 
                   
                   \path (P1ref_m1.base)+(-0.2,-4.95)  node(m2_inf) {}; 
                   
                    \path (P1ref_m1.base)+(-0.4,-4.95)  node(m3_inf) {}; 

                   \path (inf.base)+(0,1.9)  node(cont1_dec3) {}; 
                    \path (inf.base)+(0,3.9)  node(cont1_dec2) {}; 
                    \path (inf.base)+(0,5.9)  node(cont1_dec1) {}; 
                    
                   \path (inf.base)+(-2.6,0.1)  node(cont2_dec3) {};  
                   \path (inf.base)+(-2.6,2.1)  node(cont2_dec2) {}; 
                   \path (inf.base)+(-2.6,4.1)  node(cont2_dec1) {};

                    
                       \path (dec2.east)+(0,0.2)  node(P1_dec2) {};
                       \path (dec2.east)+(0,-0.2)  node(P2_dec2) {};
                       
                       \path (dec3.east)+(0,0.3)  node(P1_dec3) {};
                       \path (dec3.east)+(0,-0.3)  node(P3_dec3) {};

                        \path (dec1.east)+(1.2,0)  node(P1_hm1) {}; 
                        
                        \path (P1_dec2.base)+(1.2,0)  node(P1_hm2) {};  
                        \path (P2_dec2.base)+(1.2,0)  node(P2_hm2) {};  
                        
                        \path (dec3.east)+(1.2,0)  node(P2_hm3) {}; 
                        \path (P1_dec3.base)+(1.2,0)  node(P1_hm3) {};  
                        \path (P3_dec3.base)+(1.2,0)  node(P3_hm3) {};  
                 
                    \path (dec1.west)+(-0.3,0)  node(P1ref_y1) {}; 
                    
                     \path (dec2.west)+(0,0.15)  node(P1_ydec2) {};
                     \path (P1_ydec2.base)+(-0.3,0)  node(P1ref_y2) {};  
                     \path (dec2.west)+(-0.5,0)  node(P2ref_y2) {};

                    \path (dec3.west)+(0,0.3)  node(P1_ydec3) {};
                     \path (dec3.west)+(0,0.15)  node(P2_ydec3) {};
                     
                    \path (P1_ydec3.base)+(-0.3,0)  node(P1ref_y3) {};    
                      \path (P2_ydec3.base)+(-0.5,0)  node(P2ref_y3) {};   
                      \path (dec3.west)+(-0.7,0)  node(P3ref_y3) {};   
                      
                     \path (P1ref_y1.base)+(0,-4.95)  node(y1_inf) {}; 
                     \path (P1ref_y1.base)+(-0.2,-4.95)  node(y2_inf) {}; 
                       \path (P1ref_y1.base)+(-0.4,-4.95)  node(y3_inf) {}; 
                       
                       
                    \path (y3_inf.base)+(3.65,0)  node(fed_inf) {};
                    
                   \path (fed_inf.base)+(0,2)  node(fed_dec3) {}; 
                   \path (fed_inf.base)+(0,4)  node(fed_dec2) {}; 
                   \path (fed_inf.base)+(0,6)  node(fed_dec1) {}; 

      \path[draw ,dashed, -> , >=latex] (P1_harq1.base)-| node{}(inf.base);

      \path[draw,->,>=latex] (enc1.east)--node[shift={(0.,.2)}]{$\bx_1$}(Chan1.west);
  \path[draw,->,>=latex] (Chan1.east)--node[shift={(0,.2)}]{$\by_1$}(dec1.west);

    \path[draw,->,>=latex] (enc2.east)--node[shift={(0,.2)}]{$\bx_2$}(Chan.west);
      \path[draw,->,>=latex] (Chan.east)--node[shift={(0,.2)}]{$\by_2$}(dec2.west); 
      
        \path[draw,->,>=latex] (enc3.east)--node[shift={(0,.2)}]{$\bx_3$}(Chan3.west);
      \path[draw,->,>=latex] (Chan3.east)--node[shift={(0,.2)}]{$\by_3$}(dec3.west);

      \path[color=LHCblue,thick,draw,->,>=latex] (harq2.west)--(feed.east)node[shift={(1.051,.3)}]{$\mcF$};
       \path[color=LHCblue,thick,draw,->,>=latex] (feed.west)--node[shift={(0,0.3)}]{$\mcF$}(harq1.east);
    
         \draw [draw,->,>=latex] (dec1.east)--(P1_hm1.base)node[shift={(-0.7,0.2)}]{$\hat{\mfm}_1$};  
      \draw [draw,->,>=latex] (P1_dec2.base)--(P1_hm2.base)node[shift={(-0.7,0.2)}]{$\hat{\mfm}_1$};  
    \draw [draw,->,>=latex] (P2_dec2.base)--(P2_hm2.base)node[shift={(-0.7,0.2)}]{$\hat{\mfm}_2$};  
    
     \draw [draw,->,>=latex] (P1_dec3.base)--(P1_hm3.base)node[shift={(-0.7,0.15)}]{$\hat{\mfm}_1$};  
      \draw [draw,->,>=latex](dec3.east)--(P2_hm3.base)node[shift={(-0.7,0.15)}]{$\hat{\mfm}_2$};  
    \draw [draw,->,>=latex] (P3_dec3.base)--(P3_hm3.base)node[shift={(-0.7,0.15)}]{$\hat{\mfm}_3$};
    
  \begin{pgfonlayer}{background}       
  \draw [draw,dashed,->,>=latex] (cont1_dec1.base)--(cont2_dec1.base); 
  \draw [draw,dashed,->,>=latex] (cont1_dec2.base)--(cont2_dec2.base);  
   \draw [draw,dashed,->,>=latex] (cont1_dec3.base)--(cont2_dec3.base);  
 \end{pgfonlayer}
     

  \draw [draw,->,>=latex] (P1_m1.base)--(enc1.west)node[shift={(-0.5,0.2)}]{$\mfm_1$};    
  \draw [draw,->,>=latex] (P2_m2.base)--(P2_enc2.base)node[shift={(-1,0.2)}]{$\mfm_2$};   
  \draw [draw,->,>=latex] (P3_m3.base)--(P3_enc3.base)node[shift={(-1.2,0.2)}]{$\mfm_3$};         
       
            \path[color=LHCblue,thick,draw , <-, >=latex] (harq2.east)-| node{}(fed_inf.base) ; 
            \path[color=LHCblue,thick,draw , ->, >=latex] (dec3.north) |- node{}(fed_dec3.base) ;
            \path[color=LHCblue,thick,draw , ->, >=latex] (dec2.north) |- node{}(fed_dec2.base) ;
             \path[color=LHCblue,thick,draw , ->, >=latex] (dec1.north) |- node{}(fed_dec1.base) ;
            
      
                  \draw [fill=black] (P1ref_m1) circle (.5mm)    ;  
                  \draw [fill=black] (P2ref_m2) circle (.5mm); 
                  \draw [fill=black] (P1_m2) circle (.5mm);  
                  \draw [fill=black] (P1_m3) circle (.5mm);  
                  \draw [fill=black] (P2_m3) circle (.5mm);   
                 \draw [fill=black] (P3ref_m3) circle (.5mm); 
                    \draw [fill=black] (cont1_dec3) circle (.5mm);  
                    \draw [fill=black] (cont1_dec2) circle (.5mm); 
                    \draw [fill=black] (cont1_dec1) circle (.5mm); 

       \draw [color=black,draw,->,>=latex] (P1ref_m1.base)--(m1_inf.base);    
       \draw [color=black,draw,->,>=latex] (P2ref_m2.base)--(m2_inf.base);   
       \draw [color=black,draw,->,>=latex] (P3ref_m3.base)--(m3_inf.base);  
        
         \draw [color=black,draw,->,>=latex] (P1_m2.base)--(P1_enc2.base) ;   
        \draw [color=black,draw,->,>=latex] (P1_m3.base)--(P1_enc3.base) ;  
        \draw [color=black,draw,->,>=latex] (P2_m3.base)--(enc3.west) ;   

           \draw [color=black,draw,->,>=latex]  (P1ref_y1.base)--(y1_inf.base) ;    
           \draw [color=black,draw,->,>=latex]  (P2ref_y2.base)--(y2_inf.base) ;    
           \draw [color=black,draw,->,>=latex]  (P3ref_y3.base)--(y3_inf.base) ;    
      
           \draw [fill=black] (P1ref_y1) circle (.5mm) ; 
           \draw [fill=black] (P2ref_y2) circle (.5mm) ;     
          \draw [fill=black] (P1ref_y2) circle (.5mm) ; 
          \draw [fill=black] (P1ref_y3) circle (.5mm) ;    
          \draw [fill=black] (P2ref_y3) circle (.5mm) ; 
          \draw [fill=black] (P3ref_y3) circle (.5mm) ; 
           
              \draw [color=black,draw,->,>=latex] (P1ref_y2.base)--(P1_ydec2.base);   
             \draw [color=black,draw,->,>=latex] (P1ref_y3.base)--(P1_ydec3.base);  
             \draw [color=black,draw,->,>=latex] (P2ref_y3.base)--(P2_ydec3.base);   
          
          
          \draw [fill=black,thick] (inf.base)+(0,-.2) circle (.2mm); 
          \draw [fill=black,thick] (inf.base)+(0,-.3) circle (.2mm);
          \draw [fill=black,thick] (inf.base)+(0,-.4) circle (.2mm);

            \draw [fill=black,thick] (y2_inf.base)+(0,-.2) circle (.2mm); 
          \draw [fill=black,thick] (y2_inf.base)+(0,-.3) circle (.2mm);
          \draw [fill=black,thick] (y2_inf.base)+(0,-.4) circle (.2mm);

            \draw [fill=black,thick] (m2_inf.base)+(0,-.2) circle (.2mm); 
          \draw [fill=black,thick] (m2_inf.base)+(0,-.3) circle (.2mm);
          \draw [fill=black,thick] (m2_inf.base)+(0,-.4) circle (.2mm);

                    \draw [fill=LHCblue,color=LHCblue,thick] (fed_inf.base)+(0,-.2) circle (.2mm); 
                     \draw [fill=LHCblue,color=LHCblue,thick] (fed_inf.base)+(0,-.3) circle (.2mm);
                       \draw [fill=LHCblue,color=LHCblue,thick] (fed_inf.base)+(0,-.4) circle (.2mm);

\end{tikzpicture}

%% file: Includes/L_HARQ_optimization.tex

\section{Layer-coded HARQ}\label{sec:sbrq}

\gls{lharq} intends to remedy the difficulties steaming from the direct application of the joint coding principle. Since we cannot escape the encoding of the message $\mfm_{[k]}$ into the codeword of length $\Ns$, we will split it into simpler steps.

To understand the principle of \gls{lharq}, it is convenient to analyze a simple case of \gls{harq} with two rounds, $\kmax=2$, which we next generalize to arbitrary $\kmax$.

\subsection{The principle via example, $\kmax=2$}\label{sec:sbrq.principle}


The first transmission is done in the same way as before.  If the packet $\mfm_1$ is decoded correctly, the earned reward (normalized by $\Ns$) is given by $\mfR=R$, and a new \gls{harq} cycle starts. 

However, if the decoding fails, \ie we observe the error event, $\Err_1=\set{\hat{\mfm}_1\neq\mfm_1}$, the reward equals to $\mfR=0$ and in the second round we transmit a codeword $\bx_2$ obtained as
\begin{align}\label{codeword.2.lharq}
\bx_2&=\Phi[\mfm_{[2]}]\\
\label{mess.2.lharq}
\mfm_{[2]}&=[\mfm'_1,\mfm_2]\in\set{0,1}^{R \Ns},
\end{align}
where $\mfm_2\in\set{0,1}^{\Ns(\R-\Rs{1})}$ is a new packet and $\mfm'_1\in\set{0,1}^{\Ns\Rs{1}}$ is composed of $\Ns\Rs{1}$ bits of $\mfm_1$ (we can say that $\mfm'_1$ is a ``punctured'' version of $\mfm_1$).%

Although, per \eqref{codeword.2.lharq}, $\bx_2$ is a result of a joint encoding of packets $\mfm_1$ and $\mfm_2$, we do not decode them jointly (which would imply using $\by_1$ and $\by_2$). Instead, we decode the packet $\mfm_{[2]}$ using only the observation $\by_2$
\begin{align}\label{decoding.m.2}
\hat{\mfm}_{[2]}=\tnr{DEC}[\by_2].
\end{align}

If decoding error,  $\Err_2=\set{\hat{\mfm}_{[2]}\neq\mfm_{[2]}}$ occurs, a zero reward, $\mfR=0$, is earned and a new \gls{harq} cycle starts. However, if $\mfm_{[2]}$ is decoded correctly, we know perfectly $\mfm'_1$, see \eqref{mess.2.lharq}. Knowing these $\Ns \Rs{1}$ bits of $\mfm_{1}$, the decoder has to decode the remaining $\Ns(\R-\Rs{1})$ unknown bits using observation $\by_1$
\begin{align}\label{decoding.m.1}
\hat{\mfm}^\tnr{b}_{1}=\tnr{DEC}[\by_1;\mfm'_1],
\end{align}
where the notation $\hat{\mfm}^\tnr{b}_{1}$ is introduced to make difference with $\hat{\mfm}_1$ obtained via the direct decoding in the first round. This ``backtrack'' decoding \eqref{decoding.m.1} was introduced in~\cite{Popovski14}; a similar idea of successive decoding was also exploited in \cite{Jabi15b}.  We  define  here the backtrack decoding error by $\Errb{1}=\set{\hat{\mfm}^\tnr{b}_{1}\neq \mfm_{1}}$.  

If the decoding si successful, $\hat{\mfm}^\tnr{b}_{1}=\mfm_{1}$, the total reward is $\mfR=2\R-\Rs{1}$. Since $\Rs{1}<\R$ there is a potential for improvement over the reward $\mfR=\R$ attainable in the conventional \gls{harq}. This is because, the spirit of joint coding is followed and the second round is not merely used to convey redundancy for the packet $\mfm_1$ but also to transmit a new packet $\mfm_2$.

Let us generalize this approach.

\subsection{General case}\label{Sec:general.LHARQ}

\textbf{Encoding}

The encoding in each round is done as follows:
\begin{align}
\label{mfm.p}
\mfm'_{[l]}&=\Phi^\tnr{b}_l[\mfm_{[l]}]\in\set{0,1}^{\rho_l\Ns}\\
\label{mfm.k}
\mfm_{[k]}&=[\mfm'_{[k-1]},\mfm_k]\in\set{0,1}^{\R\Ns},\\
\label{bx.k}
\bx_k&=\Phi[\mfm_{[k]}],
\end{align}
where $\Phi^\tnr{b}_l[\cd], l=1,\ld, k-1$ are binary compressing encoders with binary rate $\R/\rho_l >1$, that is, we cannot recover $\mfm_{[l]}$  knowing solely $\mfm'_{[l]}$.
 
Since we  use the channel encoder $\Phi$ which operates with a fixed coding rate $\R$, it remains agnostic of the encoding in the step \eqref{mfm.k}; this may be contrasted with the encoding using the variable rates $\R_{[k]}$ required in the direct encoding. We thus remedied the two first difficulties related to encoding which are shown in the list in \secref{Sec:MM.HARQ}.

We introduced in \eqref{mfm.p} the notion of the compressing encoders $\Phi^\tnr{b}_k[\cd]$  to discuss the difference with \cite{Popovski14}, where the bits $\mfm'_{[k]}$ are ``parity'' bits of the packet $\mfm_{[k]}$. In many practical cases, $\Phi[\cd]$ is implemented via  \gls{bicm}, \ie it combines a binary encoder and the non-binary mapper to the symbols from the constellation $\mcX$ \cite[Sec.~2.3]{Szczecinski_Alvarado_book}. Therefore the parity bits $\mfm'_{[k]}$ might be obtained as a byproduct of the binary encoding. This also means that, as an intermediate step, the encoder $\Phi[\cd]$ must produce binary codewords longer than those necessary to produce the codewords $\bx_k$. We can thus again enter into conflict with the first item in the list of practical considerations we enumerated in \secref{Sec:MM.HARQ}. To avoid this pitfall we thus use the simplest possible compressor, that is the puncturer, \ie $\mfm'_{[k]}$ is composed of the ``systematic'' bits of $\mfm_{[k]}$.

Beside eliminating the need for the actual binary encoding by $\Phi^\tnr{b}_k[\cd]$, there are other arguments in favour of the systematic $\Phi^\tnr{b}[\cd]$ we propose. First, if the message $\mfm_{[k]}$ is successfully decoded and $\mfm_{[k-1]}$ is not, we collect the reward $\mfR=\R$, while with the parity encoding the reward would be only $\mfR=\R-\Rs{k-1}$. Second, the backtrack decoding of the message $\mfm_{[l]}$ benefits from the presence of systematic bits, more than it would from parity bits. This is particularly true for turbo-codes that we will consider, especially that current standards recommend to puncture some of the systematic bits while encoding  $\mfm_{[l]}$. These punctured bits may then be included in $\mfm'_{[l]}$ but these technical details will be discussed in~\secref{Sec:turbo.code}.%
 
\textbf{Decoding}

As for the decoding, we need of course all the observations $\by_1,\ld, \by_k$ to recover the messages $\mfm_1, \ld, \mfm_k$. However, instead of explicit joint decoding that is necessary in the direct encoding/decoding, we may use a simplified layer-by-layer decoding, defined as follows:
\begin{itemize}
\item In the $k$th round, we try to decode the packet 
\begin{align}\label{}
\hat{\mfm}_{[k]}=\tnr{DEC}[\by_k]
\end{align}
and if we succeed (\ie $\hat{\mfm}_{[k]}=\mfm_{[k]}$), we recover the message $\mfm_k$ and $\mfm'_{[k-1]}$, see \eqref{mfm.k}.
\item With $\mfm'_{[k-1]}$ at hand, we backtrack decode the packet $\mfm_{[k-1]}$
\begin{align}\label{Backtrack.decoding}
\hat{\mfm}^\tnr{b}_{[k-1]}=\tnr{DEC}[\by_{k-1}, \mfm'_{[k-1]}],
\end{align}
where we use the fact that $\mfm'_{[k-1]}$ is now known and should be used to improve the decoding results. The decoding \eqref{Backtrack.decoding} based on $\by_{k-1}$ and $\mfm'_{[k-1]}$ is stil necessary because i)~the decoding $\tnr{DEC}[\by_{k-1}]$ failed  -- that is why we are in the backtrack decoding of the $k$th round, and ii)~knowing $\mfm'_{[k-1]}$ we cannot recover $\mfm_{[k-1]}$, see the comment after \eqref{bx.k}. 
\item If there is no error, \ie $\hat{\mfm}^\tnr{b}_{[k-1]}=[\mfm'_{[k-2]},\mfm_{k-1}]$, we recover the packet $\mfm_{k-1}$ but also can go back and repeat the decoding \eqref{Backtrack.decoding} with $k\leftarrow k-1$. 
\end{itemize}

If the decoding steps are successful for $k-1, k-2, \ld, 1$ we recover all the packets $\mfm_{k-1},\ld, \mfm_1$

From the implementation point of view, the receiver operation is very simple: the decoding of $\mfm_{[k-1]}$ in \eqref{Backtrack.decoding} is done using a channel outcome $\by_{k}$ and a priori information about $\mfm_{[k-1]}$ contained in $\mfm'_{[k-1]}$. Also, the decoding result of \eqref{Backtrack.decoding}, depending  on $\SNR_{k-1}$ and $\rho_{k-1}$, is  simple to describe with the \gls{per} curves as we will  shown later. This is  very different from the decoding \eqref{joint.direct.decoding} which depends on $\SNR_1,\ld, \SNR_k$ and $\R_1, \R_{[2]}, \ld, \R_{[k]}$.

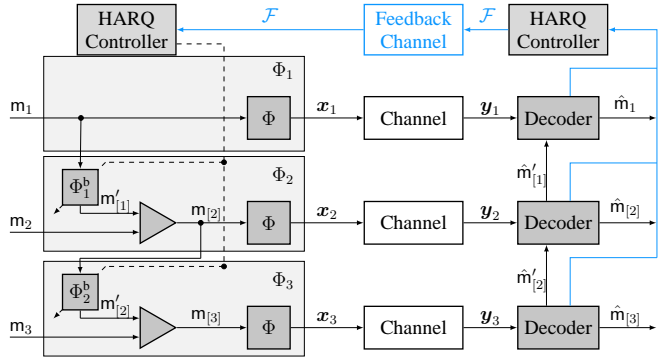
\begin{figure}[bt]
\begin{center}
\scalebox{0.63}{\input{figures/MM_model_V5}}
\end{center}
\caption{Encoding and decoding in \gls{lharq}. The \gls{harq} controller adjusts the rates of the puncturer $\Phi^\tnr{b}_k[\cd]$.}\label{Fig:L.HARQ}
\end{figure}

The two last issues from the list in \secref{Sec:MM.HARQ}, related to the decoding, are now solved. The proposed encoding/decoding schemes are illustrated in \figref{Fig:L.HARQ}, where we emphasize that the adaptation of the rate of the encoder $\Phi_k$ is done adjusting the rate of the binary compressor/puncturer $\Phi^\tnr{b}_k$.

\subsection{Throughput}\label{Sec:throughput.layered}


To calculate the throughput 
\begin{align}\label{}
\eta_{\kmax}^{\Lc}=\frac{\Ex[\mfR]}{\Ex[\mfD]}
\end{align}
we start with $\kmax=2$.

The expected reward of \gls{lharq} can  be obtained analyzing three events which produce non-zero reward:
\begin{itemize}
\item Decoding success in the first round: $\set{\ov{\Err}_1}$, where $\ov{\Err}$ denotes the complement of $\Err$;  the corresponding reward is $\mfR=\R$, 
\item Decoding success in the second round and decoding failure in the backtrack decoding: $\set{\Err_1\wedge\ov{\Err}_2\wedge \Errb{1}}$; the reward is $\mfR=\R$, and 
\item Decoding success in the second round and decoding success in the backtrack decoding: $\set{\Err_1\wedge \overline{\Err}_2 \wedge \overline{\Errb{1}}}$; the reward is $\mfR=2\R-\Rs{1}$.
\end{itemize}

The average reward can thus be calculated as
\begin{align}
\Ex[\mfR]&=\Ex\Big[\R~\IND{\overline{\Err}_1}+\R~\IND{\Err_1\wedge \overline{\Err}_2 }\nonumber \\ 
&\qquad~+(\R-\Rs{1}) \IND{\Err_1\wedge \overline{\Err}_2 \wedge \overline{\Errb{1}} } \Big]\label{eq:reaward.K.2.1}\\
&=\Ex\bigg[\R(1-\Pr\set{\Err_1})+\Pr\set{\Err_1} (1{-}\Pr\set{\Err_2})  \nonumber \\ 
&\qquad~\Big(\R+(\R-\Rs{1})\big( 1{-}{\Pr\set{\Errb{1} |\Err_1}}\big)\Big)  \bigg],\label{eq:reaward.K.2.2}
\end{align}
where $\IND{x}=1$ if $x$ is true, and $\IND{x}=0$ otherwise. The expectations in~\eqref{eq:reaward.K.2.1} are taken with respect to all variables affecting the decoding errors (including the message and the realizations of the noise), while \eqref{eq:reaward.K.2.2} takes expectation with respect to \gls{snr}s $\SNRrv_{1}, \SNRrv_{2}$.%

The expected number of transmissions is given by $\Ex[\mfD]=1+f_{1}$, where $f_1=\PR{\Err_1}$.

For~$\kmax>2$ we enumerate the decoding success/failure events in various rounds, we obtain the following generalization of~\eqref{eq:reaward.K.2.2} 
\begin{align}\label{}
\Ex[\mfR]&= \Ex\Big[\sum_{k=1}^{\kmax}(1-\Pr\set{\Err_k})\prod_{t=1}^{k-1}\Pr\set{\Err_t}\nonumber \\
& \Big(\R+\sum_{l=1}^{k-1}(\R-\Rs{l}) \prod_{z=l}^{k-1} \big( 1-\Pr\set{ \Errb{z} | \Err_z}\big) \Big)
\Big],
\end{align}
which can be expressed in a nested form as
\begin{align}\label{eq:sbrq.expec.reward}
\Ex[\mfR]= &\Ex_{\SNRrv_1}\Big[(1-\Pr\set{\Err_1})\R+\Pr\set{\Err_1} \nonumber \\
&\cd\Ex_{\SNRrv_2}\Big[(1-\Pr\set{\Err_2}) \big(\R+(\R-\Rs{1}) \nonumber \\
&\quad\cd\big( 1-\Pr\set{ \Errb{1}|\Err_1 }\big)\big)+\Pr\set{\Err_2} \nonumber \\
&\cd\Ex_{\SNRrv_3}\Big[~\ld \Big]\Big]\Big].
\end{align}

Further we note that, due to \eqref{decoding.m.2}, $\PR{\Err_{l}}$ depends only on the value of $\SNR_l$. Thus, the events $\Err_{1}, \ld, \Err_{l}$ are independent, and $f_{l}$ can be calculated as
\begin{align}\label{}
f_{l}&=\PR{ \Err_1}\ld\PR{\Err_l } 
=(f_{1})^l.
\end{align}

Thus, the average number of transmission rounds is given by
\begin{align}\label{eq:sbrq.expec.lengh}
\Ex[\mfD]&=1+f_{1}+f_{1}^2+\ld+f_{1}^{\kmax-1}
=\frac{1-f_{1}^{\kmax}}{1-f_{1}}.
\end{align}

\subsection{Optimal Rates}\label{Sec:multibit}

We are interested in finding the optimal throughput of the \gls{lharq} scheme, and we have to find the backtrack rates $\Rs{1}, \Rs{2}, \ld, \Rs{\kmax-1}$ which maximize the throughput for a given transmission rate $\R$. 

Coming back to the simple two-transmission example, the ``backtrack'' rate of the first round, $\Rs{1}\in(0,\R)$ can be defined once the decoding of $\mfm_1$ fails. Consequently, it may be adapted to the \emph{known}, but outdated, \gls{snr} $\SNR_1$. 

This idea is not new, the adaptation to the outdated channel state was already proposed in previous works, \eg \cite{Visotsky05,Szczecinski13,Pfletschinger14}, and will be exploited in~\secref{Sec:multibit} to optimize the throughput. Therefore, the rates $\Rs{k}$ are functions of \gls{snr}s $\SNR_1,\SNR_2, \ld, \SNR_{k1-}$ and eventually of other parameters defining the transmission process.%

The expected number of transmissions in~\eqref{eq:sbrq.expec.lengh} is independent of the  backtrack rates. Consequently, maximizing the throughput is equivalent to maximizing the expected reward in~\eqref{eq:sbrq.expec.reward}. Denoting its optimal value by $\overline{\mfR}$, we have
\begin{align}\label{eq:optimization.problem.sbrq}
\overline{\mfR}= &\Ex_{\SNRrv_1}\Big[\underset{\Rs{1}}{\max}~(1-\Pr\set{\Err_1})\R+\Pr\set{\Err_1}\nonumber \\
&\cd\Ex_{\SNRrv_2}\Big[\underset{\Rs{2}}{\max}~(1-\Pr\set{\Err_2}) \Big(\R+(\R-\Rs{1}) \nonumber \\
&\quad\cd\big( 1-\PR{ \Errb{1}|\Err_1 }\big)\Big)+\Pr\set{\Err_2} \nonumber \\
&\cd\Ex_{\SNRrv_3}\Big[\underset{\Rs{3}}{\max}~\ld \Big]\Big]\Big],
\end{align}
and the optimum throughput of \gls{lharq} is thus given by
\begin{align}
\eta_{\kmax}^{\Lc}=\frac{(1-f_{1}^{\kmax}) \ov{\mfR}}{1-f_{1}}.
\end{align}

The nested structure of \eqref{eq:optimization.problem.sbrq} allows us to rewrite it in the recursive form that is characteristic of~\gls{dp} in \eqref{D.P.first}--\eqref{D.P.last}, where $J_{0}\triangleq0$ and
\begin{align}\label{Jk}
J_{k}=(\R+J_{k-1}-\Rs{k})\big( 1-\PER(\SNR_k;\Rs{k}) \big)
\end{align}
has the meaning of an expected reward that may be collected thanks to the backtrack decoding.%

We also used $\PER(\SNR_k;\R)=\PR{\Err_k}$ and $\PER(\SNR;\R,\Rs{k})\triangleq\PR{\Errb{k}|\Err_{k}}$ to emphasize that the whole optimization depends solely on the \gls{per} curves of the decoder. For compactness, we define  $\PER^\tnr{c}(\cd)\triangleq 1-\PER(\cd)$.

\begin{figure*}[tb]
\begin{spacing}{1}
\begin{align}
\overline{\mfR}&= \Ex_{\SNRrv_{1}}\big[V_{1}(\SNRrv_{1},0)\big], \label{D.P.reward}\\
V_{1}(\SNR_{1},J_{0})&=\underset{\Rs{1}}{\max} \big\{ \big(\R+J_{0}\big)\PER^\tr{c}(\SNR_1;\R) +\PER(\SNR_1;\R)  \Ex_{\SNRrv_{2}}\big[V_{2}(\SNRrv_{2},J_{1})\big]\big\},   \label{D.P.first} \\ 
& \vdots   \nonumber \\    
V_{\kmax-2}(\SNR_{\kmax{-}2},J_{\kmax-3})&=\underset{\Rs{\kmax{-}2}}{\max} \big\{ \big(\R{+}J_{\kmax-3}\big)\PER^\tr{c}(\SNR_{\kmax-2};\R) +\PER(\SNR_{\kmax-2};\R)  \nonumber\\ 
&\qquad\qquad\quad \times \Ex_{\SNRrv_{\kmax-1}}\big[V_{\kmax-1}(\SNRrv_{\kmax-1},J_{\kmax-2})\big]\big\}, \\
V_{\kmax-1}(\SNR_{\kmax-1},J_{\kmax-2})&=\underset{\Rs{\kmax-1}}{\max} \big\{ \big(\R{+}J_{\kmax-2}\big)\PER^\tr{c}(\SNR_{\kmax-1};\R) +
\PER(\SNR_{\kmax-1};\R) \Ex_{\SNRrv_{\kmax}}\big[\PER^\tr{c}(\SNRrv_{\kmax};\R)\big] \nonumber\\ 
&\qquad\qquad\quad \times\big(\R+(\R+J_{\kmax-2}-\Rs{\kmax-1}) \PER^\tr{c}(\SNR_{\kmax-2};\R,\Rs{K-1}) \big)  \big\}. \label{D.P.last}
\end{align} 
\end{spacing}
\hrulefill
\end{figure*}

The optimization process starts with~\eqref{D.P.last} and continues via a backward recursion to~\eqref{D.P.reward}. In this way, thanks to the \gls{dp} formulation, the multi-dimensional global optimization in \eqref{eq:optimization.problem.sbrq} is reduced to a series of one-dimensional optimizations, and the overall computational complexity grows linearly with $\kmax$. The optimization is done point-by-point over the discretized values of the variables $(\SNR_{k}, J_{k-1})$, with $J_{k-1}\in\big(0,(k-1)\cd\R\big)$, and $\SNR_{k}\in\Real^{+}$.  In the \gls{dp} vocabulary, the variables $(\SNR_k, J_{k-1})$ form a ``state" at time $k$, the backtrack rates $\Rs{k}$ are ``actions''  and depend on the state. 

For the numerical implementation, it is convenient to truncate the \gls{per} function: we set $\PER(\SNR_k)= 0$ if $\SNR_{k}>\SNR_{\epsilon}$; where $\SNR_{\epsilon}$ satisfies $\PER(\SNR_{\epsilon})=\epsilon$. In the numerical examples, we set $\epsilon=10^{-6}$. Thus, $\Rs{k}(\SNR_{k}, J_{k-1})$ is a $2$-dimensional function, and it is non-zero only when $0 \le J_{k-1} < (k-1)\R$ and $0 \le \SNR_{k}<\SNR_{\epsilon}$.%

Since, in practice, only a limited number of rates is available, and by construction $\Rs{k} \le \R$, we use a discrete set of backtrack rates $\mcA=\{\Delta,2\Delta_{\R},\ld,\R\}$, where $\Delta=\R/\TmR$, where the number of the available rates, $\TmR$, may be adjusted to find a suitable compromise between the performance and the feedback requirements : only $\lceil \log_{2}( \TmR )\rceil$ bits of feedback are  needed even if the arguments  $(\SNR_k,J_{k-1})$ may be discretized with an arbitrary resolution when solving \eqref{D.P.first}--\eqref{D.P.last}.%

The backtrack rate functions $\Rs{k}(\SNR_{k}, J_{k-1})$ calculated off-line using \gls{dp} are stored at the receiver: after each round, the receiver observes $\SNR_k$, computes $J_{k-1}$ via \eqref{Jk}, and transmits the index of the optimal $\Rs{k}(\SNR_{k}, J_{k-1})\in\mcA$.%


%% file: figures/MM_model_V5.tex

\pgfdeclarelayer{background}
\pgfdeclarelayer{foreground}
\pgfsetlayers{background,main,foreground}

\definecolor{LHCblue}{RGB}{10, 150, 255}
\tikzstyle{form0} = [draw, minimum width=0cm, text width=0cm, 
  text centered,  minimum height=0cm]

\tikzstyle{form1} = [draw,  thick, minimum width=2cm, text width=1.8cm, fill=white, 
  text centered,  minimum height=.9cm]

\tikzstyle{form1b} = [color=LHCblue,draw,  thick, minimum width=2cm, text width=1.8cm, fill=white, 
  text centered,  minimum height=.9cm]

\tikzstyle{form1c} = [draw,  thick, minimum width=0.9cm, fill=gray!45, 
  text centered,  minimum height=.9cm]

\tikzstyle{form1d} = [draw,  thick, minimum width=2cm, text width=1.8cm, fill=gray!30, 
  text centered,  minimum height=.9cm]

\tikzstyle{form2} = [draw=none, minimum width=3.4cm, text width=1.5cm, fill=none, 
  text centered,  minimum height=0cm]    

\tikzstyle{form3} = [color=LHCblue,draw, thick,minimum width=1.6cm, text width=1.5cm, 
  text centered,  minimum height=1.6cm]
  
  \tikzstyle{form4} =[draw,  thick, minimum width=0.75cm, fill=gray!45, 
  text centered,  minimum height=0.75cm]
  
    \tikzstyle{form5} = [draw, minimum width=0.5cm, text width=0.5cm,  fill=red!20,
text centered,  minimum height=1cm]

    \tikzstyle{form7} = [draw, minimum width=0.8cm, text width=0.8cm,fill=blue!20, 
text centered,  minimum height=0.6cm]

     \tikzstyle{form6} = [draw, minimum width=2cm, text width=.4cm, 
  text centered,  minimum height=1.5cm]
  
      \tikzstyle{form8} = [draw, minimum width=1.8cm, text width=1.8cm, fill=green!50!yellow!15,
  text centered,  minimum height=2.5cm]
  
   \tikzstyle{form9} = [draw, minimum width=2.5cm, text width=3cm, fill=green!50!yellow!15,
  text centered,  minimum height=4cm]
  
    \tikzstyle{form10} = [draw, minimum width=0.8cm, text width=0.8cm,fill=none, 
text centered,  minimum height=0.6cm]

\tikzstyle{form11} = [draw, minimum width=5.5cm, text width=1.8cm, fill=gray!10, 
  text centered,  minimum height=2cm, ]
  
\tikzstyle{trigl}=[draw, thick, shape border rotate=10, isosceles triangle,isosceles triangle apex angle=60,fill=white,inner sep=0pt,minimum height=0.75cm, minimum width=0.75cm,  text centered, fill=gray!45]



\begin{tikzpicture}[trim left=-6cm]

    \node(Chan)[form1] { Channel};    
    
    \path (Chan.north)+(0,1.75)  node[form1](Chan1) {Channel}; 
     
     \path (Chan.south)+(0,-1.75)  node[form1](Chan3) {Channel}; 
            
     \path (Chan.west)+(-2,0)  node[form1c](enc2) {$\Phi$}; 
     \path (Chan.east)+(2,0)  node[form1c](dec2) {Decoder}; 
     
           \path (enc2.west)+(-2,0)  node[trigl](sum2) {}; 
            \path (sum2.west)+(-1.25,0.75)  node[form4](enc2b) {$\Phi^\tnr{b}_1$}; 
         
          \path (Chan1.west)+(-2,0)  node[form1c](enc1) {$\Phi$}; 
     \path (Chan1.east)+(2,0)  node[form1c](dec1) {Decoder}; 

        \path (Chan3.west)+(-2,0)  node[form1c](enc3) {$\Phi$}; 
     \path (Chan3.east)+(2,0)  node[form1c](dec3) {Decoder}; 
        \path (enc3.west)+(-2,0)  node[trigl](sum3) {}; 
            \path (sum3.west)+(-1.25,0.75)  node[form4](enc3b) {$\Phi^\tnr{b}_2$}; 

    \path (enc1.north)+(-3,+1.4)  node[form1d](harq1) {HARQ Controller}; 
      \path (dec1.north)+(0,+1.4)  node[form1d](harq2) {HARQ Controller}; 
      
        \path (Chan1.north)+(0,+1.4)  node[form1b](feed) {Feedback Channel}; 
        
        
         
           \path (harq1.east)+(0,-0.3)  node(P1_harq1) {};  
             \path (P1_harq1.base)+(1,0)  node(inf) {}; 
             
              \path (enc1.west)+(-5,0)  node(P1_m1) {}; 
              
               \path (enc2.west)+(0,0.2)  node(P1_enc2) {};
               \path (enc2.west)+(0,-0.2)  node(P2_enc2) {};
               
                \path (sum2.west)+(0,0.2)  node(P1_sum2) {};
               \path (sum2.west)+(0,-0.2)  node(P2_sum2) {};
                \path (P1_sum2.base)+(-1.25,0)  node(P1_sum2_encb2) {}; 
                 \path (sum2.east)+(0.5,0)  node(P1_sum2_enc2) {}; 
                 \path (sum2.east)+(0.5,-0.75)  node(P2_sum2_enc2) {}; 
                  \path (P2_sum2_enc2.base)+(-2.55,0)  node(P1_sum2_enc3b) {}; 
               
                \path (P1_enc2.base)+(-0.5,0)  node(P1_m2) {};  
                \path (P2_enc2.base)+(-5,0)  node(P2_m2) {}; 
                \path (P2_enc2.base)+(-0.7,-0)  node(P2ref_m2) {};

                 \path (enc3.west)+(0,0.3)  node(P1_enc3) {};
                 \path (enc3.west)+(0,-0.3)  node(P3_enc3) {};
                 
                 \path (sum3.west)+(0,0.2)  node(P1_sum3) {};
                 \path (sum3.west)+(0,-0.2)  node(P3_sum3) {};
                  \path (P1_sum3.base)+(-1.25,0)  node(P1_sum3_encb3) {}; 
                  
                   \path (P1_enc3.base)+(-0.5,0)  node(P1_m3) {}; 
                   \path (enc3.west)+(-0.5,0)  node(P2_m3) {};
                 \path (P3_enc3.base)+(-5,+0.1)  node(P3_m3) {}; 
                  \path (P3_enc3.base)+(-0.9,0)  node(P3ref_m3) {};

                 \path (enc1.west)+(-3.5,0)  node(P1ref_m1) {};  
                  
                   \path (P1ref_m1.base)+(0,-4.95)  node(m1_inf) {}; 
                   
                   \path (P1ref_m1.base)+(-0.2,-4.95)  node(m2_inf) {}; 
                   
                    \path (P1ref_m1.base)+(-0.4,-4.95)  node(m3_inf) {}; 

                   \path (inf.base)+(0,-2.5)  node(cont1_dec3) {}; 
                    \path (inf.base)+(0,-4.7)  node(cont1_dec2) {}; 
                    
                   \path (cont1_dec3)+(-2.5,0)  node(cont2_dec3) {};  
                   \path (cont1_dec2.base)+(-2.5,0)  node(cont2_dec2) {}; 
                   
                    \path (cont2_dec2.base)+(-1.1,-1.1)  node(cont3_dec2) {}; 
                     \path (cont2_dec3.base)+(-1.1,-1.1)  node(cont3_dec3) {};

                    
                       \path (dec2.east)+(0,0.2)  node(P1_dec2) {};
                       \path (dec2.east)+(0,-0.2)  node(P2_dec2) {};
                       
                       \path (dec3.east)+(0,0.3)  node(P1_dec3) {};
                       \path (dec3.east)+(0,-0.3)  node(P3_dec3) {};

                        \path (dec1.east)+(1.2,0)  node(P1_hm1) {}; 
                        
                        \path (dec2.east)+(1.2,0)  node(P1_hm2) {};  
                        \path (P2_dec2.base)+(1.2,0)  node(P2_hm2) {};  
                        
                        \path (dec3.east)+(1.2,0)  node(P2_hm3) {}; 
                        \path (P1_dec3.base)+(1.2,0)  node(P1_hm3) {};  
                        \path (P3_dec3.base)+(1.2,0)  node(P3_hm3) {};  
                 
                    \path (dec1.west)+(-0.3,0)  node(P1ref_y1) {}; 
                    
                     \path (dec2.west)+(0,0.15)  node(P1_ydec2) {};
                     \path (P1_ydec2.base)+(-0.3,0)  node(P1ref_y2) {};  
                     \path (dec2.west)+(-0.5,0)  node(P2ref_y2) {};

                    \path (dec3.west)+(0,0.3)  node(P1_ydec3) {};
                     \path (dec3.west)+(0,0.15)  node(P2_ydec3) {};
                     
                    \path (P1_ydec3.base)+(-0.3,0)  node(P1ref_y3) {};    
                      \path (P2_ydec3.base)+(-0.5,0)  node(P2ref_y3) {};   
                      \path (dec3.west)+(-0.7,0)  node(P3ref_y3) {};   
                      
                     \path (P1ref_y1.base)+(0,-4.95)  node(y1_inf) {}; 
                     \path (P1ref_y1.base)+(-0.2,-4.95)  node(y2_inf) {}; 
                       \path (P1ref_y1.base)+(-0.4,-4.95)  node(y3_inf) {}; 
                       
                       
                    \path (y3_inf.base)+(3.65,0)  node(fed_inf) {};
                    
                   \path (fed_inf.base)+(0,2)  node(fed_dec3) {}; 
                   \path (fed_inf.base)+(0,4)  node(fed_dec2) {}; 
                   \path (fed_inf.base)+(0,6)  node(fed_dec1) {}; 

     \path[draw ,dashed, - , >=latex] (P1_harq1.base)-| node{}(cont1_dec2.base);

      \path[draw,->,>=latex] (enc1.east)--node[shift={(0.,.2)}]{$\bx_1$}(Chan1.west);
  \path[draw,->,>=latex] (Chan1.east)--node[shift={(0,.2)}]{$\by_1$}(dec1.west);

    \path[draw,->,>=latex] (enc2.east)--node[shift={(0,.2)}]{$\bx_2$}(Chan.west);
      \path[draw,->,>=latex] (Chan.east)--node[shift={(0,.2)}]{$\by_2$}(dec2.west); 
      
        \path[draw,->,>=latex] (enc3.east)--node[shift={(0,.2)}]{$\bx_3$}(Chan3.west);
      \path[draw,->,>=latex] (Chan3.east)--node[shift={(0,.2)}]{$\by_3$}(dec3.west);

      \path[color=LHCblue,draw,->,>=latex] (harq2.west)--(feed.east)node[shift={(0.5,.3)}]{$\mcF$};
       \path[color=LHCblue,draw,->,>=latex] (feed.west)--node[shift={(0,0.3)}]{$\mcF$}(harq1.east);
    
         \draw [draw,->,>=latex] (dec1.east)--(P1_hm1.base)node[shift={(-0.65,0.3)}]{$\hat{\mfm}_1$};  
      \draw [draw,->,>=latex] (dec2.east)--(P1_hm2.base)node[shift={(-0.65,0.3)}]{$\hat{\mfm}_{[2]} $};  
    
      \draw [draw,->,>=latex](dec3.east)--(P2_hm3.base)node[shift={(-0.65,0.25)}]{$\hat{\mfm}_{[3]}$};  
    
     \begin{pgfonlayer}{background}       
     \path (Chan.west)+(-4,0.4)  node[form11](adap2) {}; 
     \path (Chan3.west)+(-4,0.4)  node[form11](adap1) {}; 
     \path (Chan1.west)+(-4,0.3)  node[form11](adap1) {}; 
     \path (Chan.west)+(-1.7,1)  node() {$\Phi_2$};  
      \path (Chan1.west)+(-1.7,1)  node() {$\Phi_1$};  
       \path (Chan3.west)+(-1.7,1)  node() {$\Phi_3$};  
     
 \end{pgfonlayer}
    
  \begin{pgfonlayer}{background}       
  \draw [draw,dashed,-,>=latex] (cont1_dec2.base)--(cont2_dec2.base);  
   \draw [draw,dashed,->,>=latex] (cont2_dec2.base)--(cont3_dec2.base);  
   
   \draw [draw,dashed,-,>=latex] (cont1_dec3.base)--(cont2_dec3.base);  
    \draw [draw,dashed,->,>=latex] (cont2_dec3.base)--(cont3_dec3.base);  
 \end{pgfonlayer}
     

  \draw [draw,->,>=latex] (P1_m1.base)--(enc1.west)node[shift={(-4.75,0.2)}]{$\mfm_1$};    
  \draw [draw,->,>=latex] (P2_m2.base)--(P2_sum2.base)node[shift={(-2.5,0.2)}]{$\mfm_2$};   
  \draw [draw,->,>=latex] (P3_m3.base)--(P3_sum3.base)node[shift={(-2.5,0.2)}]{$\mfm_3$};         
       
            \path (dec3.north)+(0.25,0)  node(P_dec3) {}; 
            \path (dec2.north)+(0.25,0)  node(P_dec2) {}; 
            \path (dec1.north)+(0.25,0)  node(P_dec1) {}; 
            
           \path[color=LHCblue,draw , <-, >=latex] (harq2.east)-| node{}(fed_dec3.base) ; 
            \path[color=LHCblue,draw , -, >=latex] (P_dec3.base) |- node{}(fed_dec3.base) ;
            \path[color=LHCblue,draw , -, >=latex] (P_dec2.base) |- node{}(fed_dec2.base) ;
            \path[color=LHCblue,draw , -, >=latex] (P_dec1.base) |- node{}(fed_dec1.base) ;
            \path (dec3.north)+(-0.25,0)  node(P1_dec3) {}; 
            \path (dec2.north)+(-0.25,0)  node(P1_dec2) {}; 
            \path (dec1.south)+(-0.25,0)  node(P2_dec1) {}; 
             \path (dec2.south)+(-0.25,0)  node(P2_dec2) {}; 
             \draw [draw,->,>=latex] (P1_dec2.base)--(P2_dec1.base)node[shift={(-0.3,-0.75)}]{$\hat{\mfm}'_{[1]}$}; 
             \draw [draw,->,>=latex] (P1_dec3.base)--(P2_dec2.base)node[shift={(-0.3,-0.75)}]{$\hat{\mfm}'_{[2]}$}; 
             
      
                 \draw [fill=black] (P1ref_m1) circle (.5mm)    ;  
                    \draw [fill=black] (cont1_dec3) circle (.5mm);  
                    \draw [fill=black] (cont1_dec2) circle (.5mm); 

       \draw [color=black,draw,->,>=latex] (P1ref_m1.base)--(enc2b.north);    
       \draw [color=black,draw,-,>=latex] (enc2b.south)--(P1_sum2_encb2.base);   
       
       \draw [color=black,draw,->,>=latex] (P1_sum2_encb2.base)--(P1_sum2.base)node[shift={(-0.5,0.25)}]{$\mfm'_{[1]}$};    
        \draw [color=black,draw,->,>=latex] (sum2.east)--(enc2.west)node[shift={(-0.85,0.2)}]{$\mfm_{[2]}$};    
        
 \draw [color=black,draw,-,>=latex] (enc3b.south)--(P1_sum3_encb3.base);   
  \draw [color=black,draw,->,>=latex] (P1_sum3_encb3.base)--(P1_sum3.base)node[shift={(-0.5,0.25)}]{$\mfm'_{[2]}$};    
   \draw [color=black,draw,->,>=latex] (sum3.east)--(enc3.west)node[shift={(-0.85,0.2)}]{$\mfm_{[3]}$};    
   
    \draw [color=black,draw,-,>=latex] (P1_sum2_enc2.base)--(P2_sum2_enc2.base);   
    \draw [color=black,draw,-,>=latex] (P2_sum2_enc2.base)--(P1_sum2_enc3b.base);   
    \draw [color=black,draw,->,>=latex] (P1_sum2_enc3b.base)--(enc3b.north);   
     \draw [fill=black] (P1_sum2_enc2) circle (.5mm);

\end{tikzpicture}

%% file: Includes/Numerical.tex
\section{Numerical examples}\label{Sec:Numerical}
Numerical results illustrating the optimization procedure explained in~\secref{Sec:multibit} are here shown in two cases. First, we will use synthetic decoder curves which will allow the reader to reproduce the results. Next, we will use experimental \gls{per} curves obtained using turbo-codes to show the throughput gains in a realistic scenario and shed some light on the practical aspects of the encoding.

\subsection{Synthetic PER curves}\label{Sec:Synthetic}

We will use the well-known model for the \gls{per} curve \cite{Liu04}
\begin{align}
\label{PER.SNR}
\PER(\SNR,\R)&=
\begin{cases}
1 &\text{if}\quad \SNR<\SNR_{\tr{th}}\\
\exp\big({-}\tilde{a} (\SNR/\SNR_{\tr{th}}-1) \big) &\text{if}\quad \SNR\geq\SNR_{\tr{th}}
\end{cases};
\end{align}
where $I(\SNR_{\tr{th}})=\R$ and $I(x)=\log_2(1+x)$; as indicated in \cite{Sassioui16}, $\tilde{a}=4$ may be fitted to empirical curves.

To characterize the decoding errors in~\gls{irharq}, we use the simplified approach proposed in~\cite{Pauli2007,Wan06}, where we apply the \gls{per} curve~\eqref{PER.SNR} 
\begin{align}\label{eq:per.irahrq}
\Pr\set{\Err_k}\approx\PER(\SNR^\Sigma_k,\R),
\end{align}
and use the \emph{aggregate} \gls{snr}  given by
\begin{align}\label{aggreg.SNR}
\SNR^\Sigma_k=I^{-1}\Big( \sum_{l=1}^k I(\SNR_l)\Big).
\end{align}

Note that, setting $\tilde{a}=\infty$, we conveniently fall back on the idealized threshold decoding of~\cite{Caire01,Larsson14,Jabi16}. 

Regarding \gls{lharq}, we need to characterize the decoder \gls{per} curve in the backtrack decoding. Since the effective rate of the message is decreased, we use 
\begin{align}\label{eq:per.sbrq}
\Pr\set{\Errb{k}}=\PER(\SNR_{k};\R-\Rs{k}).
\end{align}

From the assumption of backward errors implication~\cite{Gu06,Long09}, $\Errb{k}\Rightarrow \Err_{k}$
 (which means that if the decoding fails in the backtrack phase, it must have failed in the original transmission), we have 
\begin{align}\label{eq:per.back.sbrq}
\PR{\Err_{k}\wedge \Errb{k}}&\approx\PR{\Errb{k}},\\
\PR{\Errb{k}|\Err_k}&\approx\frac{\PER(\SNR_{k};\R-\Rs{k})}{\PER(\SNR_{k};\R)}.
\end{align}

Furthermore, with the backward errors implication assumption, $\Err_k\Rightarrow\Err_{k-1}\Rightarrow \ld\Rightarrow\Err_1$, $f_{k}$ is calculated as
\begin{align}\label{eq:fk.IRHARQ}
f_{k}\approx \Ex\big[\PER(\SNRrv^\Sigma_k,\R)\big],
\end{align}
where the expectation is taken over the channel \gls{snr}s which contribute to $\SNRrv^\Sigma_k$ via \eqref{aggreg.SNR}.

\begin{figure}[tb]
\psfrag{I}[ct][ct][\siz]{$I$}
\psfrag{Parity 3}[cb][cc][\siz]{$\Rs{3}(\SNR_{3},J_{2})$}
\psfrag{Parity 2}[cb][cc][\siz]{$\Rs{2}(\SNR_{2},J_{1})$}
\psfrag{Parity 1}[cb][cc][\siz]{$\Rs{1}(\SNR_{1},0)$}
\psfrag{SNR3}[cb][cc][\siz]{$\SNR_{3}$\dB}
\psfrag{SNR2}[cb][cc][\siz]{$\SNR_{2}$\dB}
\psfrag{SNR1}[cb][cc][\siz]{$\SNR_{1}\dB$}
\psfrag{Reward2}[cb][cc][\siz]{$J_{2}$}
\psfrag{Reward1}[cb][cc][\siz]{$J_{1}$}
\begin{center}
\scalebox{\sizf}{\includegraphics[width=\linewidth]{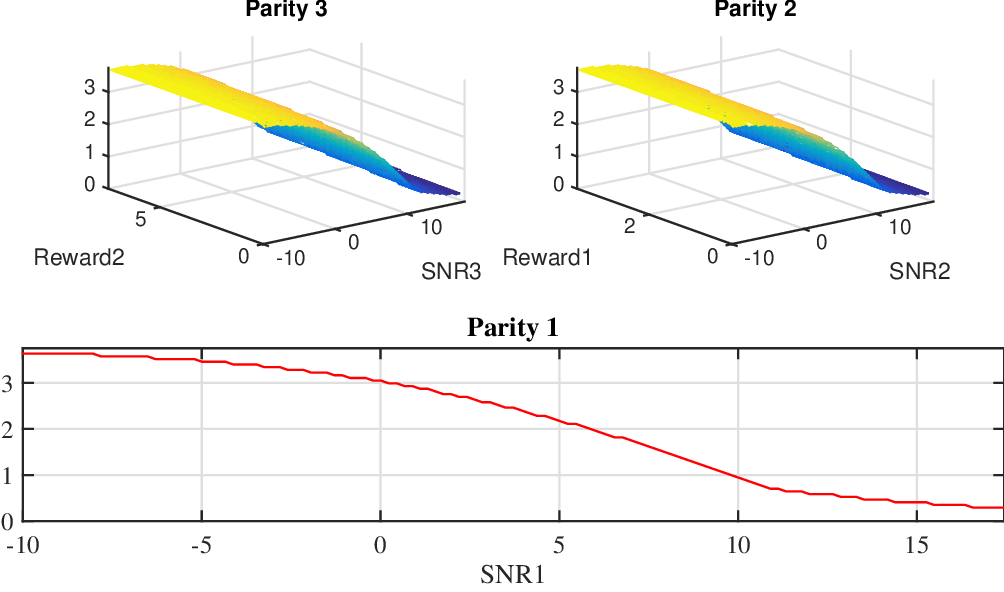}}
\caption{\gls{lharq} optimal policies $\Rs{k}(\SNR_{k},J_{k-1})$ obtained for $\R=3.75$, $\kmax=4$, $\SNRav=15\dBval$, and the synthetic \gls{per} curves defined in \secref{Sec:Synthetic}.}\label{Fig:Rayleigh.SBRQ.policy}
\end{center}
\end{figure}

The optimal backtrack rates, $\Rs{k}(\SNR_{k}, J_{k-1})$, obtained with the \gls{dp} formulation are shown in~\figref{Fig:Rayleigh.SBRQ.policy}. The rates $\Rs{k}$ decrease with the observed $\SNR_{k}$ because they are optimized to increase the chances of success in the backtrack decoding, and yet not to penalize the throughput. Thus, as $\SNR_{k}$ increases, the number of bits needed to \emph{guarantee} the backtrack decoding decreases. We also observe that the optimal policy varies little in terms $J_{k-1}$, which indicates the possibility of using a suboptimal policy independent of $J_{k-1}$ as we will discuss in~\secref{Sec:heuristic}.%

\begin{figure}[tb]
\input{./figures/Fig.SBRQ.ANSBRQ.IRHARQ.Gaussian.Modulation.tex}
\caption{Throughput of the proposed \gls{lharq}, $\eta_{\kmax}^{\Lc}$, is compared to the throughput of \gls{irharq}, $\eta^{\IR}_K$; $\R=3.75$, $\log_{2}(\TmR)=6$, and the synthetic \gls{per} curves defined in \secref{Sec:Synthetic}.}\label{Fig:Rayleigh.SBRQ.ANSBRQ}
\end{figure}
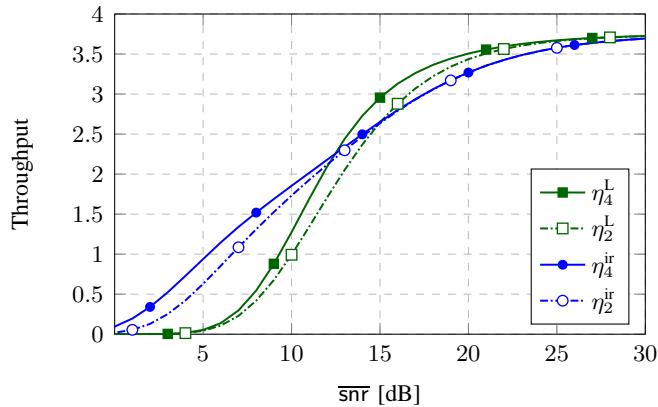

The throughputs of \gls{lharq} and  \gls{irharq} are compared in~\figref{Fig:Rayleigh.SBRQ.ANSBRQ}. As already mentioned in~\secref{Sec:MM.HARQ}, we are mostly interested in the throughput close to $\R$ where the conventional \gls{irharq} fails to provide gains even when increasing the number of retransmissions \cite{Larsson14}. Indeed, this is where the improvement from \gls{lharq} materializes. For instance, around a throughput of $\eta=3$, \gls{lharq} offers a gain of approximately $1$~\!dB compared to \gls{irharq} with $\kmax=2$, and up to $2.5$dB with $\kmax=4$. On the other hand, \gls{lharq} is outperformed by \gls{irharq} for small values of the throughput, where $f_{1}$ is high. This is not  a serious drawback because, knowing the average \gls{snr}, we may switch to \gls{irharq} if necessary or, if possible, use a different rate $\R$. Performing a joint decoding, \ie decoding $\mfm_{[2]}$ from $\by_2$ and $\by_1$ would also improve the performance at the cost of increased complexity, as we discussed in~\secref{Sec:MM.HARQ}.%

\begin{figure}[tb]
\input{./figures/Fig.SBRQ.Discetization.effect.Gaussian.Modulation.tex}
\caption{Throughput of the \gls{lharq}, $\eta_{\kmax}^{\Lc}$, is compared to the throughput of \gls{irharq}, $\eta^{\IR}_K$, for $\R=3.75$ and different numbers of feedback bits $\log_{2}( \TmR )$. The synthetic \gls{per} curves defined in \secref{Sec:Synthetic} are used.}\label{Fig:Rayleigh.SBRQ.Discetization}
\end{figure}
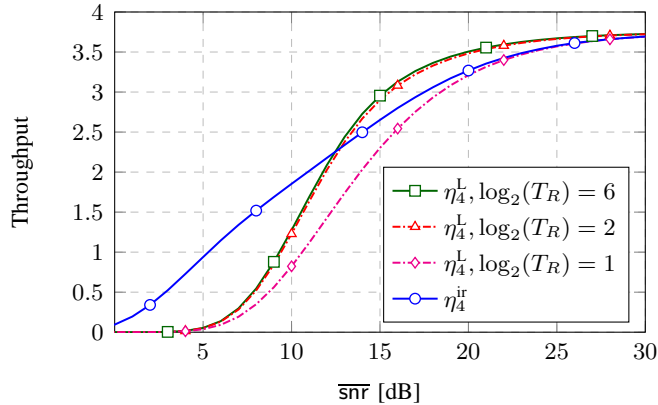

Finally, \figref{Fig:Rayleigh.SBRQ.Discetization} provides an insight into the additional feedback required to make \gls{lharq} operational. We note that with only two additional feedback bits, \gls{lharq} practically attains its maximum potential and ensures notable gains over the conventional \gls{irharq}.

\subsection{Rate Adaptation with Turbo-Codes}\label{Sec:turbo.code}

\begin{figure}[t] 
\begin{center}
(a)
\\
\scalebox{1}{\input{./figures/Fig.WEP3.3gpp.tex}}
\\
(b)
\\
\scalebox{1}{\input{./figures/Fig.WEP.Rate2.25.3gpp.tex}}
\caption{$\Pr\{\Err_{k}\wedge \Errb{k}\}$ as a function of the instantaneous $\SNR_{k}$ for different values of $\Rs{k}$ when a turbo-code and a $16$-QAM modulation are used with (a) $\R=3.75$ and (b) $\R=2.25$. Dashed curves correspond to the case where systematic bits are not punctured, \ie $\R_{\mfm}=0\%$, while solid lines correspond to the results obtained by puncturing systematic bits with $\R_{\mfm}=6.25\%$. }\label{Fig:Rayleigh.WEP}
\end{center}
\end{figure}
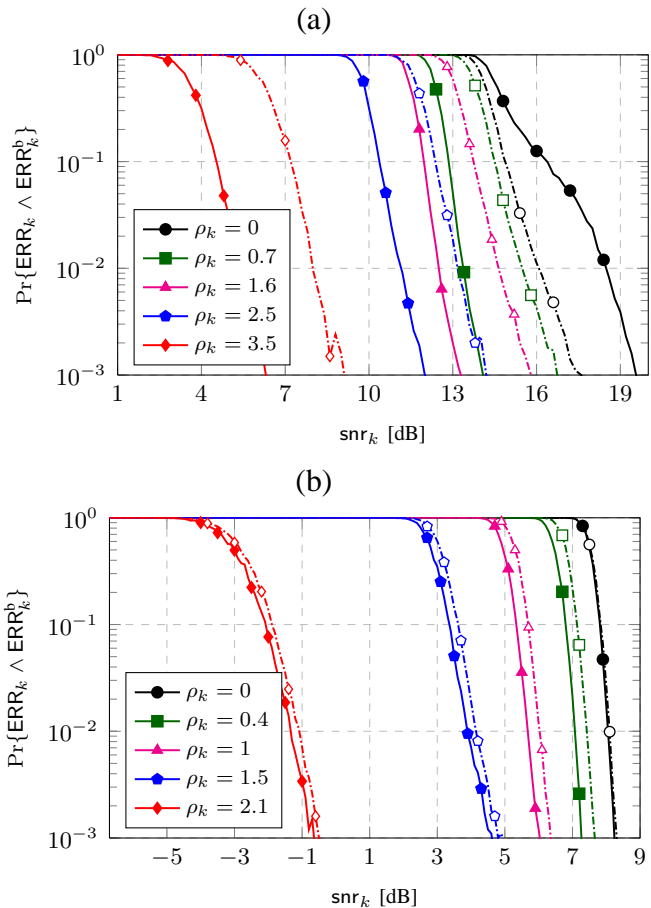

In order to perform the optimization steps \eqref{D.P.first}--\eqref{D.P.last} for practical encoders/decoders, we only need the \gls{per} curves $\PER(\SNR;\R)$ and $\PER(\SNR; \R, \Rs{})$. These are obtained by simulating/measuring $\PR{\Err_{k}}$ and $\Pr\set{\Err_{k}\wedge \Errb{k}}$, and  the results obtained for different values of $\Rs{k}$ are shown in~\figref{Fig:Rayleigh.WEP}; of course, if $\Rs{k}=0$ we have $\PR{\Err_{k}}=\Pr\set{\Err_{k}\wedge \Errb{k}}$. 


We used here a turbo-code specified by \gls{3gpp} in \cite{3gpp36.212}, comprising two constituent convolutional encoders with generating polynomials $[13/15]_8$ and the \gls{3gpp} pseudo-random interleaver defined in \cite[Sec. 5.1.3.2.3]{3gpp36.212}. The result of the encoding, after the interleaving of subblocks as prescribed by the \gls{3gpp} rate matching algorithm \cite[Sec. 5.1.4.1]{3gpp36.212} is denoted by $\mfc = [\mfm,\mfm^\tnr{p}]$, where $\mfm^\tnr{p}$ and $\mfm$ are interleaved versions of the parity bits and systematic bits, respectively. 

Since we use $\R\in\set{2.25, 3.75}$ and the nominal coding rate of the \gls{3gpp} encoder is $r_\tnr{o}=1/3$, we need to puncture $\mfc$ to obtain the binary coding rate $r=\R/m\in\set{0.5625, 0.9375}$, where $m=4$ is the rate of the 16-\gls{qam} modulation. We thus take $\Nc'=r_\tnr{o}\Nc/r$ bits from $\mfc$ and map them with a Gray mapping  \cite[Sec.~2.5.2]{Szczecinski_Alvarado_book} onto $\Ns=1024$ symbols $\bx_k$ taken from a 16-\gls{qam} constellation, which are next transmitted over the channel~\eqref{y.x.z}.  The receiver calculates the \glspl{llr} using exact expressions \cite[Sec.~3.3]{Szczecinski_Alvarado_book} and feeds them to the \gls{bcjr} decoder \cite{BCJR} implemented in the log-domain; the interested reader can refer to the library at~\cite{Doray15}.

As for the puncturing, we take $\Nc'$ bits starting with the offset of $\R_{\mfm}~[\%]$ defining the percentage of the systematic bits being punctured. In this way, the codeword $\bx_k$ in the $k$th round contains $100\%-\R_{\mfm}$ of the bits in the message $\mfm_{[k]}$. The interesting question now is: which bits $\mfm'_{[k]}$ from the message $\mfm_{[k]}$ should be taken to construct the message $\mfm_{[k+1]}=[\mfm'_{[k]},\mfm_{k+1}]$?

The interplay between the coding and the \gls{harq} scheme becomes, indeed, interesting: for $\R_\mfm>0$, it is beneficial to construct $\mfm'_{[k]}$ using the \emph{first} bits of $\mfm_{[k]}$ because some of these bits are punctured to construct $\bx_k$ in round $k$; thus, knowing these bits (after a successful decoding in round $k+1$) improves the performance of the decoder in the backtrack phase. On the other hand, if we construct $\mfm'_{[k]}$ using the \emph{last} bits of $\mfm_{[k]}$, their perfect knowledge (after a successful decoding of $\mfm_{[k+1]}$) will eliminate the channel-related \glspl{llr} during the backtrack decoding, removing thus some of the available information.

We show the \gls{per} curves of the turbo-decoder in~\figref{Fig:Rayleigh.WEP} for $\R_\mfm=0\%$ and $\R_\mfm=6.25\%$, where the latter offset value is, in fact, recommended by  the \gls{3gpp}. The important observation is that while the results of $\PER(\SNR_k;\R)$ (circles) deteriorate due to the puncturing of the systematic bits (solid lines, $\R_\mfm=6.25\%$), the results of the backtrack decoding are significantly improved in this case. There is thus a tradeoff between decreasing the decoding error probability and decreasing the probability of backtrack decoding error $\Pr\set{\Errb{k}\wedge\Err_k}$. This tradeoff becomes even clearer as the nominal transmission rate $\R$ increases. 

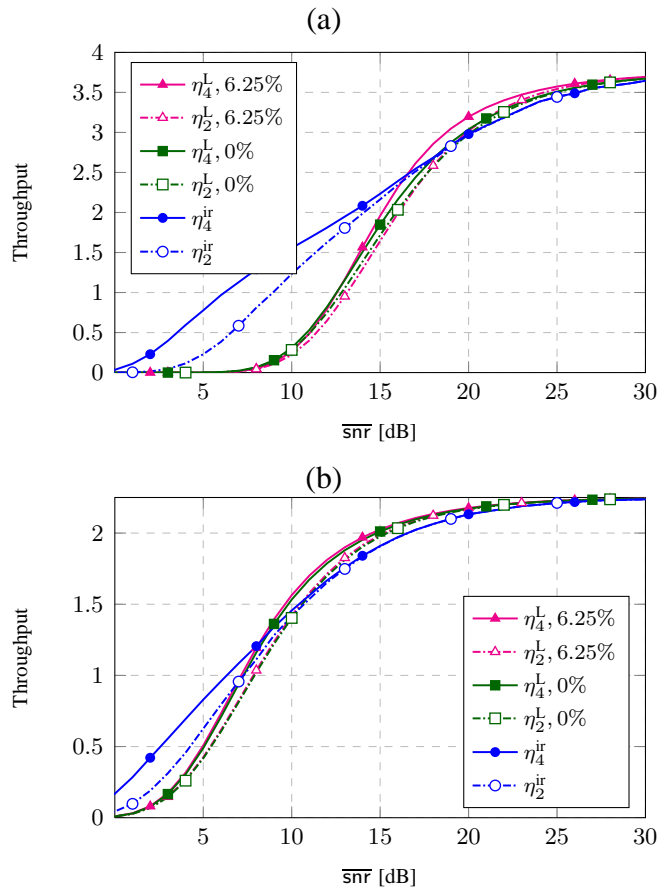
\begin{figure}[t] 
\begin{center}
(a)
\\
\scalebox{1}{\input{./figures/Fig.SBRQ.IRHARQ.Turbo.Code3.3gpp.tex}}
\\
(b)
\\
\scalebox{1}{\input{./figures/Fig.SBRQ.IRHARQ.Turbo.Code.Rate.2.25.3gpp.tex}}
\caption{The throughput of \gls{lharq} and \gls{irharq} obtained for turbo-coded $16$QAM transmissions with the puncturing defined by $\R_{\mfm}=0\%$ and $\R_{\mfm}=6.25\%$for (a) $\R=3.75$, and (b) $\R=2.25$; $\log_{2}( \TmR )=4$. }\label{Fig:Turbo.Rayleigh.SBRQ}
\end{center}
\end{figure}

The above mentioned tradeoff becomes evident with the throughput results shown in~\figref{Fig:Turbo.Rayleigh.SBRQ} based on the same turbo-code \gls{per} curves shown in~\figref{Fig:Rayleigh.WEP}. For $\R=3.75$, and using $\R_{\mfm}=6\%$, the gain of \gls{lharq} over \gls{irharq} is $\sim0.5$~\!dB for $\kmax=2$, and $\sim2.5$~\!dB for $\kmax=4$ (measured at $\eta=3$). On the other hand, a similar gain is obtained for $\kmax=2$ with $\R_{\mfm}=0\%$, but no further improvement is observed when the number of transmissions is increased to $\kmax=4$. However, the effect of changing $\R_{\mfm}$ on the results of \gls{lharq} is less notable when $\R=2.25$ as can be seen in~\figref{Fig:Turbo.Rayleigh.SBRQ}(b). This is not too surprising, since the difference between $\Pr\{\Err_{k}\wedge \Errb{k}\}$ curves of $\R_{\mfm}=0\%$ and $\R_{\mfm}=6.25\%$ is less important when $\R=2.25$; see \figref{Fig:Rayleigh.WEP}b.

%% file: figures/Fig.SBRQ.ANSBRQ.IRHARQ.Gaussian.Modulation.tex
%

\centering
\newcommand{\lsize}{\footnotesize}
\pgfplotsset{tick label style={font=\footnotesize},
    label style={font=\footnotesize},
    legend style={font=\footnotesize}
}
\tikzset{every mark/.append style={mark size = 2, solid, fill=white, line width=0.5pt,}}
\begin{tikzpicture}
 \begin{axis}[
	axis lines = box,
	xlabel={\lsize $\SNRav$\dB},
	ylabel={\lsize  Throughput},
	xmin=0,xmax=30,
	ymin=0,ymax=4,
	xtick={5,10,...,30},
	ytick={0,0.5,..., 4},
	legend pos=south east,
	xmajorgrids=true,
	ymajorgrids=true,
	grid style=dashed,
	clip=true,
	legend style={
		draw=black,
		cells={anchor=west},
	},
]

\addplot[color=black!60!green, line width=0.75pt,mark=square*,mark repeat=6,mark phase=4,mark options={scale=0.8, fill=black!60!green}]
table[x=SNRdB, y=Throughput2]{./figures/Gaussian.SBRQ.Rayleigh.dat};
\addlegendentry{$\eta_4^{\Lc}$}
\addplot[color=black!60!green, line width=0.75pt,densely dashdotted,mark=square*,mark repeat=6,mark phase=5]
table[x=SNRdB, y=Throughput1]{./figures/Gaussian.SBRQ.Rayleigh.dat};
\addlegendentry{$\eta_2^{\Lc}$}
\addplot[color=blue , line width=0.75pt,mark=*,mark repeat=6,mark phase=3,mark options={scale=0.75, fill=blue}]
table[x=SNRdB, y=Throughput2]{./figures/Gaussian.C.HARQ.Rayleigh.dat};
\addlegendentry{$\eta_4^{\IR}$}
\addplot[color=blue , line width=0.75pt,densely dashdotted,mark=*,mark repeat=6,mark phase=2,]
table[x=SNRdB, y=Throughput1]{./figures/Gaussian.C.HARQ.Rayleigh.dat};
\addlegendentry{$\eta_2^{\IR}$}


\end{axis}

\end{tikzpicture}

%% file: figures/Fig.SBRQ.Discetization.effect.Gaussian.Modulation.tex
%

\centering
\newcommand{\lsize}{\footnotesize}
\pgfplotsset{tick label style={font=\footnotesize},
    label style={font=\footnotesize},
    legend style={font=\footnotesize}
}
\tikzset{every mark/.append style={mark size = 2, solid, fill=white, line width=0.5pt,}}
\begin{tikzpicture}
 \begin{axis}[
	axis lines = box,
	xlabel={\lsize $\SNRav$\dB},
	ylabel={\lsize  Throughput},
	xmin=0,xmax=30,
	ymin=0,ymax=4,
	xtick={5,10,...,30},
	ytick={0,0.5,..., 4},
	legend pos=south east,
	xmajorgrids=true,
	ymajorgrids=true,
	grid style=dashed,
	clip=true,
	legend style={
		draw=black,
		cells={anchor=west},
	},
]

\addplot[color=black!60!green, line width=0.75pt,mark=square*,mark repeat=6,mark phase=4]
table[x=SNRdB, y=Throughput5]{./figures/Gaussian.SBRQ.HARQ.Rayleigh.Discretization.effect.dat};
\addlegendentry{$\eta_4^{\Lc}, \log_{2}( \TmR )=6$}
\addplot[color=red, line width=0.75pt,densely dashdotted,mark=triangle*,mark repeat=6,mark phase=5]
table[x=SNRdB, y=Throughput2]{./figures/Gaussian.SBRQ.HARQ.Rayleigh.Discretization.effect.dat};
\addlegendentry{$\eta_4^{\Lc}, \log_{2}( \TmR )=2$}
\addplot[color=magenta, line width=0.75pt,densely dashdotted,mark=diamond*,mark repeat=6,mark phase=5]
table[x=SNRdB, y=Throughput1]{./figures/Gaussian.SBRQ.HARQ.Rayleigh.Discretization.effect.dat};
\addlegendentry{$\eta_4^{\Lc}, \log_{2}( \TmR )=1$}
\addplot[color=blue , line width=0.75pt,mark=*,mark repeat=6,mark phase=3]
table[x=SNRdB, y=Throughput2]{./figures/Gaussian.C.HARQ.Rayleigh.dat};
\addlegendentry{$\eta_4^{\IR}$}


\end{axis}

\end{tikzpicture}

%% file: figures/Fig.WEP3.3gpp.tex
%

\centering
\newcommand{\lsize}{\scriptsize}
\pgfplotsset{tick label style={font=\footnotesize},
    label style={font=\footnotesize},
    legend style={font=\footnotesize},
}
\tikzset{every mark/.append style={mark size = 2, solid, fill=white, line width=0.5pt,}}
\begin{tikzpicture}
 \begin{axis}[
	axis lines = box,
	xlabel={\lsize $\SNR_{k}$ [dB]},
	ylabel={\lsize  $\Pr\{\Err_{k}\wedge \Errb{k}\}$},
	ymode = log,
	xmin=1,xmax=20,
	ymin=1e-3,ymax=1,
	xtick={1,4,...,20},
	legend pos=south west,
	xmajorgrids=true,
	ymajorgrids=true,
	grid style=dashed,
	clip=true,
	legend style={
		draw=black,
		cells={anchor=west},
	},
]

\addplot[color=black, line width=0.75pt,mark=*, mark color=black, mark repeat=6,mark phase=70,mark options={fill=black}]
table[x=SNR, y=WEP1]{./figures/WEP.Rayleigh.poincenage6.3gpp.dat};
\addlegendentry{\lsize $\Rs{k}=0$}
\addplot[color=black!60!green, line width=0.75pt,mark=square*,mark repeat=5,mark phase=58,mark options={fill=black!60!green}]
table[x=SNR, y=WEP2]{./figures/WEP.Rayleigh.poincenage6.3gpp.dat};
\addlegendentry{\lsize $\Rs{k}=0.7$}
\addplot[color=magenta, line width=0.75pt,mark=triangle*,mark repeat=4,mark phase=55,mark options={fill=magenta}]
table[x=SNR, y=WEP3]{./figures/WEP.Rayleigh.poincenage6.3gpp.dat};
\addlegendentry{\lsize $\Rs{k}=1.6$}
\addplot[color=blue, line width=0.75pt,mark=pentagon*,mark repeat=4,mark phase=45,mark options={fill=blue}]
table[x=SNR, y=WEP4]{./figures/WEP.Rayleigh.poincenage6.3gpp.dat};
\addlegendentry{\lsize $\Rs{k}=2.5$}
\addplot[color=red, line width=0.75pt,mark=diamond*,mark repeat=5,mark phase=10,mark options={fill=red}]
table[x=SNR, y=WEP5]{./figures/WEP.Rayleigh.poincenage6.3gpp.dat};
\addlegendentry{\lsize $\Rs{k}=3.5$}

\addplot[color=black, line width=0.75pt,densely dashdotted,mark=*, mark repeat=6,mark phase=73]
table[x=SNR, y=WEP1]{./figures/WEP.Rayleigh.poincenage0.3gpp.dat};
\addplot[color=black!60!green, line width=0.75pt,densely dashdotted,mark=square*,mark repeat=5,mark phase=65]
table[x=SNR, y=WEP2]{./figures/WEP.Rayleigh.poincenage0.3gpp.dat};
\addplot[color=magenta, line width=0.75pt,densely dashdotted,mark=triangle*,mark repeat=4,mark phase=60]
table[x=SNR, y=WEP3]{./figures/WEP.Rayleigh.poincenage0.3gpp.dat};
\addplot[color=blue, line width=0.75pt,densely dashdotted,mark=pentagon*,mark repeat=5,mark phase=55,]
table[x=SNR, y=WEP4]{./figures/WEP.Rayleigh.poincenage0.3gpp.dat};
\addplot[color=red, line width=0.75pt,densely dashdotted,mark=diamond*,mark repeat=8,mark phase=23]
table[x=SNR, y=WEP5]{./figures/WEP.Rayleigh.poincenage0.3gpp.dat};

\end{axis}

\end{tikzpicture}

%% file: figures/Fig.WEP.Rate2.25.3gpp.tex
%

\centering
\newcommand{\lsize}{\scriptsize}
\pgfplotsset{tick label style={font=\footnotesize},
    label style={font=\footnotesize},
    legend style={font=\footnotesize},
}
\tikzset{every mark/.append style={mark size = 2, solid, fill=white, line width=0.5pt,}}
\begin{tikzpicture}
 \begin{axis}[
	axis lines = box,
	xlabel={\lsize $\SNR_{k}$ [dB]},
	ylabel={\lsize  $\Pr\{\Err_{k}\wedge \Errb{k}\}$},
	ymode = log,
	xmin=-6.7,xmax=9,
	ymin=1e-3,ymax=1,
	xtick={-7,-5,...,9},
	legend pos=south west,
	xmajorgrids=true,
	ymajorgrids=true,
	grid style=dashed,
	clip=true,
	legend style={
		draw=black,
		cells={anchor=west},
	},
]

\addplot[color=black, line width=0.75pt,mark=*, mark color=black, mark repeat=6,mark phase=141,mark options={fill=black}]
table[x=SNR, y=WEP1]{./figures/WEP.Rayleigh.poincenage6.Rate2.25.3gpp.dat};
\addlegendentry{\lsize $\Rs{k}=0$}
\addplot[color=black!60!green, line width=0.75pt,mark=square*,mark repeat=5,mark phase=135,mark options={fill=black!60!green}]
table[x=SNR, y=WEP2]{./figures/WEP.Rayleigh.poincenage6.Rate2.25.3gpp.dat};
\addlegendentry{\lsize $\Rs{k}=0.4$}
\addplot[color=magenta, line width=0.75pt,mark=triangle*,mark repeat=4,mark phase=115,mark options={fill=magenta}]
table[x=SNR, y=WEP3]{./figures/WEP.Rayleigh.poincenage6.Rate2.25.3gpp.dat};
\addlegendentry{\lsize $\Rs{k}=1$}
\addplot[color=blue, line width=0.75pt,mark=pentagon*,mark repeat=4,mark phase=95,mark options={fill=blue}]
table[x=SNR, y=WEP4]{./figures/WEP.Rayleigh.poincenage6.Rate2.25.3gpp.dat};
\addlegendentry{\lsize $\Rs{k}=1.5$}
\addplot[color=red, line width=0.75pt,mark=diamond*,mark repeat=5,mark phase=28,mark options={fill=red}]
table[x=SNR, y=WEP5]{./figures/WEP.Rayleigh.poincenage6.Rate2.25.3gpp.dat};
\addlegendentry{\lsize $\Rs{k}=2.1$}

\addplot[color=black, line width=0.75pt,densely dashdotted,mark=*, mark repeat=6,mark phase=143]
table[x=SNR, y=WEP1]{./figures/WEP.Rayleigh.poincenage0.Rate2.25.3gpp.dat};
\addplot[color=black!60!green, line width=0.75pt,densely dashdotted,mark=square*,mark repeat=5,mark phase=135]
table[x=SNR, y=WEP2]{./figures/WEP.Rayleigh.poincenage0.Rate2.25.3gpp.dat};
\addplot[color=magenta, line width=0.75pt,densely dashdotted,mark=triangle*,mark repeat=4,mark phase=117]
table[x=SNR, y=WEP3]{./figures/WEP.Rayleigh.poincenage0.Rate2.25.3gpp.dat};
\addplot[color=blue, line width=0.75pt,densely dashdotted,mark=pentagon*,mark repeat=5,mark phase=95,]
table[x=SNR, y=WEP4]{./figures/WEP.Rayleigh.poincenage0.Rate2.25.3gpp.dat};
\addplot[color=red, line width=0.75pt,densely dashdotted,mark=diamond*,mark repeat=8,mark phase=30]
table[x=SNR, y=WEP5]{./figures/WEP.Rayleigh.poincenage0.Rate2.25.3gpp.dat};

\end{axis}

\end{tikzpicture}

%% file: figures/Fig.SBRQ.IRHARQ.Turbo.Code3.3gpp.tex
%

\centering
\newcommand{\lsize}{\scriptsize}
\pgfplotsset{tick label style={font=\footnotesize},
    label style={font=\footnotesize},
    legend style={font=\footnotesize}
}
\tikzset{every mark/.append style={mark size = 2, solid, fill=white, line width=0.5pt,}}
\begin{tikzpicture}
 \begin{axis}[
	axis lines = box,
	xlabel={\lsize $\SNRav$\dB},
	ylabel={\lsize  Throughput},
	xmin=0,xmax=30,
	ymin=0,ymax=4,
	xtick={5,10,...,30},
	ytick={0,0.5,..., 4},
	legend pos=north west,
	xmajorgrids=true,
	ymajorgrids=true,
	grid style=dashed,
	clip=true,
	legend style={
		draw=black,
		cells={anchor=west},
	},
]

\addplot[color=magenta, line width=0.75pt,mark=triangle*,mark repeat=6,mark phase=3,mark options={scale=0.85, fill=magenta}]
table[x=SNRdB, y=Throughput2]{./figures/Turbo.code.SBRQ.Rayleigh.poincenage6.3gpp.dat};
\addlegendentry{\lsize $\eta_4^{\Lc}, 6.25 \%$}
\addplot[color=magenta, line width=0.75pt,densely dashdotted,mark=triangle*,mark repeat=5,mark phase=4,]
table[x=SNRdB, y=Throughput1]{./figures/Turbo.code.SBRQ.Rayleigh.poincenage6.3gpp.dat};
\addlegendentry{\lsize $\eta_2^{\Lc}, 6.25 \%$}

\addplot[color=black!60!green, line width=0.75pt,mark=square*,mark repeat=6,mark phase=4,mark options={scale=0.8, fill=black!60!green}]
table[x=SNRdB, y=Throughput2]{./figures/Turbo.code.SBRQ.Rayleigh.poincenage0.3gpp.dat};
\addlegendentry{\lsize $\eta_4^{\Lc}, 0 \%$}
\addplot[color=black!60!green, line width=0.75pt,densely dashdotted,mark=square*,mark repeat=6,mark phase=5]
table[x=SNRdB, y=Throughput1]{./figures/Turbo.code.SBRQ.Rayleigh.poincenage0.3gpp.dat};
\addlegendentry{\lsize $\eta_2^{\Lc}, 0 \%$}
\addplot[color=blue , line width=0.75pt,mark=*,mark repeat=6,mark phase=3,mark options={scale=0.75, fill=blue}]
table[x=SNRdB, y=Throughput2]{./figures/Turbo.Code.C.HARQ.Rayleigh.3gpp.2.dat};
\addlegendentry{\lsize $\eta_4^{\IR}$}
\addplot[color=blue , line width=0.75pt,densely dashdotted,mark=*,mark repeat=6,mark phase=2,]
table[x=SNRdB, y=Throughput1]{./figures/Turbo.Code.C.HARQ.Rayleigh.3gpp.2.dat};
\addlegendentry{\lsize $\eta_2^{\IR}$}


\end{axis}

\end{tikzpicture}

%% file: figures/Fig.SBRQ.IRHARQ.Turbo.Code.Rate.2.25.3gpp.tex
%

\centering
\newcommand{\lsize}{\scriptsize}
\pgfplotsset{tick label style={font=\footnotesize},
    label style={font=\footnotesize},
    legend style={font=\footnotesize}
}
\tikzset{every mark/.append style={mark size = 2, solid, fill=white, line width=0.5pt,}}
\begin{tikzpicture}
 \begin{axis}[
	axis lines = box,
	xlabel={\lsize $\SNRav$\dB},
	ylabel={\lsize  Throughput},
	xmin=0,xmax=30,
	ymin=0,ymax=2.25,
	xtick={5,10,...,30},
	ytick={0,0.5,..., 4},
	legend pos=south east,
	xmajorgrids=true,
	ymajorgrids=true,
	grid style=dashed,
	clip=true,
	legend style={
		draw=black,
		cells={anchor=west},
	},
]

\addplot[color=magenta, line width=0.75pt,mark=triangle*,mark repeat=6,mark phase=3,mark options={scale=0.85, fill=magenta}]
table[x=SNRdB, y=Throughput2]{./figures/Turbo.code.SBRQ.Rayleigh.poincenage6.Rate2.25.3gpp.dat};
\addlegendentry{\lsize $\eta_4^{\Lc}, 6.25 \%$}
\addplot[color=magenta, line width=0.75pt,densely dashdotted,mark=triangle*,mark repeat=5,mark phase=4,]
table[x=SNRdB, y=Throughput1]{./figures/Turbo.code.SBRQ.Rayleigh.poincenage6.Rate2.25.3gpp.dat};
\addlegendentry{\lsize $\eta_2^{\Lc}, 6.25 \%$}

\addplot[color=black!60!green, line width=0.75pt,mark=square*,mark repeat=6,mark phase=4,mark options={scale=0.8, fill=black!60!green}]
table[x=SNRdB, y=Throughput2]{./figures/Turbo.code.SBRQ.Rayleigh.poincenage0.Rate2.25.3gpp.dat};
\addlegendentry{\lsize $\eta_4^{\Lc}, 0 \%$}
\addplot[color=black!60!green, line width=0.75pt,densely dashdotted,mark=square*,mark repeat=6,mark phase=5]
table[x=SNRdB, y=Throughput1]{./figures/Turbo.code.SBRQ.Rayleigh.poincenage0.Rate2.25.3gpp.dat};
\addlegendentry{\lsize $\eta_2^{\Lc}, 0 \%$}
\addplot[color=blue , line width=0.75pt,mark=*,mark repeat=6,mark phase=3,mark options={scale=0.75, fill=blue}]
table[x=SNRdB, y=Throughput2]{./figures/Turbo.Code.C.HARQ.Rayleigh.Rate2.25.3gpp.dat};
\addlegendentry{\lsize $\eta_4^{\IR}$}
\addplot[color=blue , line width=0.75pt,densely dashdotted,mark=*,mark repeat=6,mark phase=2,]
table[x=SNRdB, y=Throughput1]{./figures/Turbo.Code.C.HARQ.Rayleigh.Rate2.25.3gpp.dat};
\addlegendentry{\lsize $\eta_2^{\IR}$}


\end{axis}

\end{tikzpicture}

%% file: Includes/Discussion.tex
\section{Sub-optimal rate adaptation policies}\label{sec:discussion}

We will now discuss adaptation strategies aiming  i)~to streamline the way the backtrack errors are handled, and ii)~to simplify the rates adaptation.

\subsection{All-or-none decoding}\label{Sec:AoN}

\begin{figure*}[tb]
\begin{spacing}{1}
\begin{align}
\overline{\mfR}^{\AN}&=\R\cd \Ex_{\SNRrv_{1}}\big[V_{1}(\SNRrv_{1},R)\big], \label{D.P.reward.AN}\\
V_{1}(\SNR_{1},J_{0})&=\underset{\Rs{1}}{\max} \Big\{\PER^\tr{c}(\SNR_1;\R)+\frac{J_{0}+\R-\Rs{1}}{J_{0}}\PER(\SNR_1;\R)\PER^\tr{c}(\SNR_1;\R,\Rs{1}) \nonumber\\
&\qquad\qquad\quad \times\Ex_{\SNRrv_{2}}\big[V_{2}(\SNRrv_{2},J_{1})\big] \Big\}, \label{D.P.first.AN} \\ 
& \vdots   \nonumber \\    
V_{\kmax-2}(\SNR_{\kmax-2},J_{\kmax-3})&=\underset{\Rs{\kmax-2}}{\max} \Big\{\PER^\tr{c}(\SNR_{\kmax-2};\R)+\frac{J_{\kmax-3}+\R-\Rs{\kmax-2}}{J_{\kmax-3}}\PER(\SNR_{\kmax-2};\R)\PER^\tr{c}(\SNR_{\kmax-2};\R,\Rs{\kmax-2}) \nonumber\\
&\qquad\qquad\quad \times\Ex_{\SNRrv_{\kmax-1}}\big[V_{\kmax-1}(\SNRrv_{\kmax-1},J_{\kmax-2})\big] \Big\},\\
V_{\kmax-1}(\SNR_{\kmax-1},J_{\kmax-2})&=\underset{\Rs{\kmax-1}}{\max} \Big\{\PER^\tr{c}(\SNR_{\kmax-1};\R)+\frac{J_{\kmax-2}+\R-\Rs{\kmax-1}}{J_{\kmax-2}}\PER(\SNR_{\kmax-1};\R)\PER^\tr{c}(\SNR_{\kmax-1};\R,\Rs{\kmax-1})\nonumber\\
&\qquad\qquad\quad\times\Ex_{\SNRrv_{\kmax}}\big[\PER^\tr{c}(\SNRrv_{\kmax};\R)\big] \Big\}. \label{D.P.last.AN}
\end{align} 
\end{spacing}
\hrulefill
\end{figure*}

In the example of two rounds, presented in~\secref{sec:sbrq.principle}, if the message $\mfm_{[2]}$ is decoded successfully and the backtrack decoding of $\mfm_{1}$ fails, \gls{lharq} does not discard the correctly received $\Ns \Rs{1}$ bits of $\mfm_{1}$ (meaning that only a part of $\mfm_{1}$ is received correctly). This complicates the buffer management, and may not be suitable for some applications in which only the packet $\mfm_{1}$ is critical and the packets $\mfm_2, \ld, \mfm_k$ are piggybacked on the ongoing \gls{harq} process to not waste the ressources.

We thus want to evaluate a different strategy, where a non-zero reward is collected only if both $\mfm_{[2]}$ and $\mfm_{1}$ are decoded successfully. In the resulting \gls{anlharq} the average reward \eqref{eq:reaward.K.2.1} is modified as 
\begin{align}
\Ex[\mfR]&=\Ex\Big[\R~\IND{\overline{\Err}_1}+(2\R-\Rs{1})~\IND{\Err_1\wedge \overline{\Err}_2 \wedge \overline{\Errb{1}} } \Big]\nonumber \\
&=\Ex\bigg[\R(1-\Pr\set{\Err_1})+(2\R-\Rs{1}) (1{-}\Pr\set{\Err_2})  \nonumber \\ 
&\qquad~\Pr\set{\Err_1} \big( 1{-}{\Pr\set{\Errb{1} |\Err_1}}\big)  \bigg].\label{eq:reward.anlharq.K.2}
\end{align}

In a case of arbitrary $\kmax$ the expected reward of \gls{anlharq} \eqref{eq:reward.anlharq.K.2} generalizes as follows:
\begin{align}
\Ex[\mfR]&=\Ex\Big[\sum_{k=1}^{\kmax}(k\R-\sum_{l=1}^{k-1}\Rs{l}) \cd \big(1-\Pr\set{\Err_k}) \nonumber \\
&\qquad\times\prod_{z=1}^{k-1} \Pr\set{\Err_z} \big( 1-\Pr\set{\Errb{z} | \Err_z}\big) \Big],\label{eq:ansbrq.expec.reward} \\
&=\R~\Ex_{\SNRrv_1}\bigg[(1-\Pr\set{\Err_1})+\frac{(2\R-\Rs{1})}{\R} \Pr\set{\Err_1} \nonumber \\ 
&\qquad~ \big(1{-}{\Pr\set{\Errb{1} |\Err_1}}\big) \Ex_{\SNRrv_2} \Big[ \big( 1{-}\Pr\set{\Err_2}\big)+\nonumber\\
&\qquad~ \frac{(3\R-\Rs{1}-\Rs{2})}{(2\R-\Rs{1})} \Pr\set{\Err_2}\big(1{-}{\Pr\set{\Errb{2} |\Err_2}}\big) \nonumber \\ 
&\qquad~  \Ex_{\SNRrv_3} \Big[ \big( 1{-}\Pr\set{\Err_3})+\ld \Big]\Big]\Big],\label{eq:reward.anlharq.K}
\end{align}
while the expected number of rounds is the same as in~\eqref{eq:sbrq.expec.lengh}. Thus, the optimal throughput of \gls{anlharq}, denoted as $\eta_{\kmax}^{\AN}$, is given by
\begin{align}
\eta_{\kmax}^{\AN}=\frac{(1-f_{1}^{\kmax})\cd \overline{\mfR}^{\AN}}{1-f_{1}},
\end{align}
where $\overline{\mfR}^{\AN}$ denotes the optimum expected reward \eqref{eq:ansbrq.expec.reward} with respect to $\left\{\Rs{k}\right\}_{k=1}^{\kmax-1}$. Again, profiting from the nested structure of \eqref{eq:reward.anlharq.K}, the $\overline{\mfR}^{\AN}$ can be found by solving the recursive equations \eqref{D.P.reward.AN}--\eqref{D.P.last.AN}, where $J_{k}\in\big(\R,(k+1)\cd\R\big)$, and it is related to $J_{k-1}$ and $\Rs{k}$ through
\begin{align}
J_{k}=J_{k-1}+\R-\Rs{k},
\end{align}
where, by definition, $J_{0}=\R$.

\begin{figure}[tb] 
\input{./figures/Fig.SBRQ.ANSBRQ.IRHARQ.Turbo.Code3.3gpp.tex}
\caption{The throughputs of \gls{anlharq} and the heuristic policy \eqref{eq:heuristic_policy} when $\epsilon=0.1$ are compared with \gls{lharq} results obtained for turbo-coded $16$QAM transmissions with the puncturing defined by $\R_{\mfm}=6.25\%$ for $\R=3.75$; $\log_{2}( \TmR )=4$. }\label{Fig:Turbo.Rayleigh.SBRQ.ANSBRQ}
\end{figure}
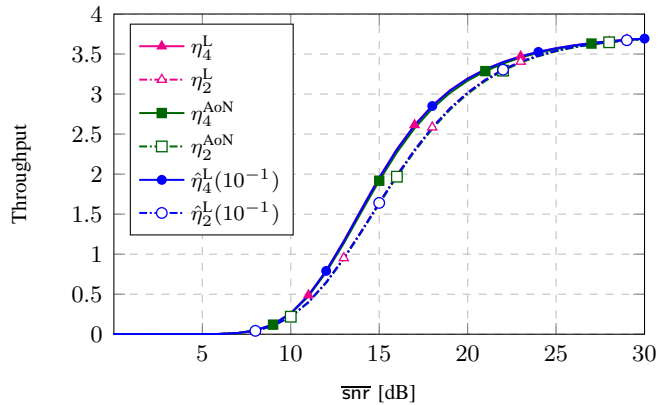

The results of the proposed \gls{anlharq} are compared with \gls{lharq} in~\figref{Fig:Turbo.Rayleigh.SBRQ.ANSBRQ}. We can clearly see that imposing the constraint that all backtrack decoding actions are successful does not penalize the final throughput of \gls{anlharq}, which is practically equal to the optimal throughput of \gls{lharq}. We thus conclude that the optimal backtrack rates of \gls{lharq} are such to guarantee a high probability of successful backtrack decoding. This observation will be exploited in the following to simplify the rate adaptation policy.

\subsection{Fixed-outage policy} \label{Sec:heuristic}

The rate adaptation policies $\Rs{k}(\SNR_{k},J_{k-1})$ determined by solving \eqref{D.P.first}--\eqref{D.P.last} or \eqref{D.P.first.AN}--\eqref{D.P.last.AN} are sufficient to optimize the throughput but they have two drawbacks, namely
\begin{enumerate}
\item The rates are three-dimensional functions of $\SNR_k$, $J_{k-1}$ and the transmission round $k$, see \figref{Fig:Rayleigh.SBRQ.policy}; this is inconvenient from the point of view of storage requirement.
\item The rate  depend on the distribution of $\SNRrv$, which not only adds to the storage and optimization complexity, but makes the solution potentially sensitive to  the changes in the channel model.
\end{enumerate}

To address the above issues, we propose a simple one-dimensional adaptation policy, independent of $J_{k-1}$, $k$, and $\pdf_{\SNRrv}(\SNR)$, which is partially inspired by  the form of the optimal policy in~\figref{Fig:Rayleigh.SBRQ.policy} that  varies little in terms of $J_{k-1}$ and $k$.  Moreover, motivated by the results of \gls{anlharq}, which provide results with very reliable backtrack decoding and this, without penalizing the throughput, we propose the rate adaptation policy, which will guarantee successful instantaneous backtrack decoding. Thus we take into account solely the outdated channel \gls{snr}
\begin{align}\label{eq:heuristic_policy}
\Rs{}(\SNR_{k})=\argmin_{\Rs{} \in \mcA}\big\{\Rs{}~|~\PER(\SNR_{k};\R,\Rs{}) \le \epsilon \big\},
\end{align}
where $\epsilon \in \Real_+$ is a design parameter. 

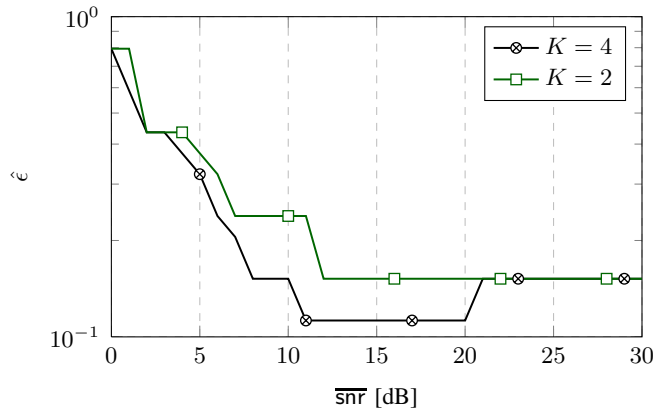
\begin{figure}[tb] 
\input{./figures/Fig.epsilon.optimal.heuristic.tex}
\caption{The optimal $\hat{\epsilon}$ which solves \eqref{eta:heuristic_policy} for turbo-coded $16$QAM transmissions with the puncturing defined by $\R_{\mfm}=6.25\%$ for $\R=3.75$; $\log_{2}( \TmR )=4$.}\label{Fig:Turbo.Rayleigh.optimal.epsilon}
\end{figure}

The throughput obtained with the policy $\Rs{}(\SNR_{k})$, we denote by $\hat{\eta}_{\kmax}^{\Lc}(\epsilon)$, can be evaluated via \eqref{eq:sbrq.expec.reward} to determine the optimal values of $\epsilon$ 
\begin{align}\label{eta:heuristic_policy}
\hat{\epsilon}=\argmax_{\epsilon}~\hat{\eta}_{\kmax}^{\Lc}(\epsilon)
\end{align}
which we show in~\figref{Fig:Turbo.Rayleigh.optimal.epsilon}. Alternatively, we might use simulations to evaluate the throughput with different values of $\epsilon$; the direct advantage of such an approach is that it would free us from the channel-model dependence.
 
Here, we observe that while $\hat{\epsilon}$ is a function of the average \gls{snr}, it varies little in the region of  high $\SNRav$. And since this region of operation is of main interest, we further fix $\epsilon=10^{-1}$ eliminating the dependence of the policy on the channel statistics.\footnote{This value is arbitrary, but we wanted a ``round" number close to what the results indicated.} The throughput $\hat{\eta}_{\kmax}^{\Lc}(10^{-1})$ is shown in \figref{Fig:Turbo.Rayleigh.SBRQ.ANSBRQ}, where it is clear that the penalty  incurred with respect to the optimal solution is negligible. 

This is quite a remarquable result which indicates that  the throughput obtained with a very simple adaptation strategy \eqref{eq:heuristic_policy} that is agnostic to the channel statistics as well as to the past and the future of the \gls{harq} process, is very close to the optimal solution.

%% file: figures/Fig.SBRQ.ANSBRQ.IRHARQ.Turbo.Code3.3gpp.tex
%

\centering
\newcommand{\lsize}{\scriptsize}
\pgfplotsset{tick label style={font=\footnotesize},
    label style={font=\footnotesize},
    legend style={font=\footnotesize}
}
\tikzset{every mark/.append style={mark size = 2, solid, fill=white, line width=0.5pt,}}
\begin{tikzpicture}
 \begin{axis}[
	axis lines = box,
	xlabel={\lsize $\SNRav$\dB},
	ylabel={\lsize  Throughput},
	xmin=0,xmax=30,
	ymin=0,ymax=4,
	xtick={5,10,...,30},
	ytick={0,0.5,..., 4},
	legend pos=north west,
	xmajorgrids=true,
	ymajorgrids=true,
	grid style=dashed,
	clip=true,
	legend style={
		draw=black,
		cells={anchor=west},
	},
]

\addplot[color=magenta, line width=0.75pt,mark=triangle*,mark repeat=6,mark phase=12,mark options={scale=0.85, fill=magenta}]
table[x=SNRdB, y=Throughput2]{./figures/Turbo.code.SBRQ.Rayleigh.poincenage6.3gpp.dat};
\addlegendentry{\lsize $\eta_4^{\Lc}$}
\addplot[color=magenta, line width=0.75pt,densely dashdotted,mark=triangle*,mark repeat=5,mark phase=9,]
table[x=SNRdB, y=Throughput1]{./figures/Turbo.code.SBRQ.Rayleigh.poincenage6.3gpp.dat};
\addlegendentry{\lsize $\eta_2^{\Lc}$}
\addplot[color=black!60!green, line width=0.75pt,mark=square*,mark repeat=6,mark phase=10,mark options={scale=0.8, fill=black!60!green}]
table[x=SNRdB, y=Throughput2]{./figures/Turbo.code.ANSBRQ.Rayleigh.3gpp.dat};
\addlegendentry{\lsize $\eta_4^{\AN}$}
\addplot[color=black!60!green, line width=0.75pt,densely dashdotted,mark=square*,mark repeat=6,mark phase=11]
table[x=SNRdB, y=Throughput1]{./figures/Turbo.code.ANSBRQ.Rayleigh.3gpp.dat};
\addlegendentry{\lsize $\eta_2^{\AN}$}
\addplot[color=blue , line width=0.75pt,mark=*,mark repeat=6,mark phase=13,mark options={scale=0.75, fill=blue}]
table[x=SNRdB, y=Throughput2]{./figures/Turbo.code.heuristic.Rayleigh.3gpp.epsilon0.1.dat};
\addlegendentry{\lsize $\hat{\eta}_4^{\Lc}(10^{-1})$}
\addplot[color=blue , line width=0.75pt,densely dashdotted,mark=*,mark repeat=7,mark phase=9,]
table[x=SNRdB, y=Throughput1]{./figures/Turbo.code.heuristic.Rayleigh.3gpp.epsilon0.1.dat};
\addlegendentry{\lsize $\hat{\eta}_2^{\Lc}(10^{-1})$}


\end{axis}

\end{tikzpicture}

%% file: figures/Fig.epsilon.optimal.heuristic.tex
%

\centering
\newcommand{\lsize}{\footnotesize}
\pgfplotsset{tick label style={font=\footnotesize},
    label style={font=\footnotesize},
    legend style={font=\footnotesize},
}
\tikzset{every mark/.append style={mark size = 2, solid, fill=white, line width=0.5pt,}}
\begin{tikzpicture}
 \begin{axis}[
	axis lines = box,
	xlabel={\lsize $\SNRav$\dB},
	ylabel={\lsize  $\hat{\epsilon}$},
	ymode = log,
	xmin=0,xmax=30,
	ymin=1e-1,ymax=1,
	xtick={0,5,...,30},
	legend pos=north east,
	xmajorgrids=true,
	ymajorgrids=true,
	grid style=dashed,
	clip=true,
	legend style={
		draw=black,
		cells={anchor=west},
	},
]

\addplot[color=black, line width=0.75pt,mark=otimes*, mark repeat=6,mark phase=6]
table[x=SNRdB, y=outage2]{./figures/Turbo.code.heuristic.outage.Rayleigh.3gpp.dat};
\addlegendentry{$\kmax=4$}
\addplot[color=black!60!green, line width=0.75pt,mark=square*,mark repeat=6,mark phase=5]
table[x=SNRdB, y=outage1]{./figures/Turbo.code.heuristic.outage.Rayleigh.3gpp.dat};
\addlegendentry{$\kmax=2$}

%
\end{axis}

\end{tikzpicture}